\documentclass[aps,pra,twocolumn,floatfix,nofootinbib,showpacs,superscriptaddress]{revtex4-2}
\usepackage{mathbbol}
\usepackage{amsmath}
\usepackage{tikz}
\usepackage{centernot}

\usepackage[utf8]{inputenc}
\usepackage[english]{babel}

\usepackage{amsfonts}
\usepackage{amsthm}
\usepackage{mathrsfs}
\usepackage{mathtools} 

\usepackage{tabularx}
\usepackage{physics,subfigure}

\usepackage{hyperref}

\begin{document}
\title{Perturbed graphs achieve unit transport efficiency without environmental noise}
\author{Simone Cavazzoni}
\email{simone.cavazzoni@unimore.it}
\affiliation{Dipartimento di Scienze Fisiche, Informatiche e Matematiche, Universit\`{a} di Modena e Reggio Emilia, I-41125 Modena, Italy}
\author{Luca Razzoli}
\email{luca.razzoli@uninsubria.it}
\affiliation{Dipartimento di Scienze Fisiche, Informatiche e Matematiche, Universit\`{a} di Modena e Reggio Emilia, I-41125 Modena, Italy}
\affiliation{Dipartimento di Scienza e Alta Tecnologia, Universit\`a degli Studi dell’Insubria, I-22100 Como, Italy}
\author{Paolo Bordone}
\email{paolo.bordone@unimore.it}
\affiliation{Dipartimento di Scienze Fisiche, Informatiche e Matematiche, Universit\`{a} di Modena e Reggio Emilia, I-41125 Modena, Italy}
\affiliation{Centro S3, CNR-Istituto di Nanoscienze, I-41125 Modena, Italy}
\author{Matteo G. A. Paris}
\email{matteo.paris@fisica.unimi.it}
\affiliation{Quantum Technology Lab, Dipartimento di Fisica {\em Aldo Pontremoli}, Universit\`{a} degli Studi di Milano, I-20133 Milano, Italy}
\affiliation{INFN, Sezione di Milano, I-20133 Milano, Italy}
\date{\today}
\begin{abstract}
Coherent transport of an excitation through a network corresponds to continuous-time quantum walk on a graph, and the transport properties of the system may be radically different depending on the graph and on the initial state. The transport efficiency, i.e., the integrated probability of trapping at a certain vertex, is a measure of the success rate of the transfer process. Purely coherent quantum transport is known to be less efficient than the observed excitation transport, e.g., in biological systems, and there is evidence that environmental noise is indeed crucial for excitation transport. At variance with this picture, we here address purely coherent transport on highly symmetric graphs, and show analytically that it is possible to enhance the transport efficiency without environmental noise, i.e., using only a minimal perturbation of the graph. In particular, we show that 
adding an extra weight to one or two edges, depending on whether the initial state is localized or in a superposition of two vertex states, breaks the inherent symmetries of the graph and may be sufficient
to achieve unit transport efficiency. We also briefly discuss the conditions to obtain a null transport efficiency,
i.e., to avoid trapping.
\end{abstract}
\maketitle
\section{Introduction}
\label{Introduction}
A continuous-time quantum walk (CTQW) describes the dynamics of a quantum particle evolving continuously in time in a discrete space according to the  Schr\"odinger equation \cite{portugal2018quantum}. CTQWs were introduced in the study of decision trees for computational problems, when Farhi and Gutmann basically proposed
to promote the classically constructed transition rate matrix to Hamiltonian of a quantum system \cite{farhi1998quantum}.
This is why, nowadays, the Laplacian matrix, which mathematically represents a graph and is a suitable generator of continuous-time classical random walks, is taken as Hamiltonian of the prototypical CTQW. However, any Hermitian operator which respects the topology of the graph is a proper CTQW Hamiltonian \cite{dalessandro2010connection}.

In recent years, CTQWs have attracted increasing attention being a suitable toy model whose inherent quantum nature determines its main features, e.g., the self-interference effects and generation of coherence, and which intrinsically depends on the given discrete topology. CTQWs have different physical implementations \cite{wang2013physical} and a wide range of application, from quantum computation \cite{childs2009universal,lahini2018quantum,childs2004spatiaL,chakraborty2020optimality,campos2021quantum,tamascelli2014quantum} to quantum communication \cite{christandl2005perfect,kendon2011perfect,alvir2016perfect}, and from modeling of physical phenomena \cite{lahini2012quantum} to neural networks \cite{schuld2014quantum,dalla2020quantumstate}, and complex networks \cite{tsomokos2011quantum,faccin2014community,moutinho2021quantum}.

A relevant application of CTQWs lies in the description of coherent transport of an excitation through a network \cite{mulken2011continuous,blumen2006coherent,mulken2007quantum,xu2008coherent,agliari2008dynamics,rai2008transport,salimi2010continuous,darazs2014transport,yalouz2018continuous,li2020quantum}, which itself serves to benchmark CTQWs against their classical analog in terms of  transport efficiency, whose measure, e.g., can rely on the density of states \cite{mulken2006efficiency}. The present work fits into the context of {excitation transfer} with trapping \cite{mulken2007survival,agliari2010continuous}, and therefore a proper measure of transport efficiency is the overall probability of trapping at a certain vertex \cite{olaya2008efficiency,rebentrost2009environment}. Biological systems, {noisy by nature}, are known to show quantum effects \cite{engel2007evidence,collini2010coherently,lambert2013quantum,mohseni2014quantum} and transport processes whose efficiency is higher than what would be observed in either the purely quantum or purely classical cases. There is evidence that environmental noise is crucial for efficient excitation transport, in biological  \cite{olaya2008efficiency,mohseni2008environment,plenio2008dephasing,caruso2009highly,rebentrost2009environment,hoyer2010limits,kassal2012environment} and nonbiological systems \cite{leon2015noise,biggerstaff2016enhancing,maier2019environment,kurt2020efficient,chisholm2021stochastic}.

Despite this observation, and to properly understand the role of coherence and quantumness in transport, we here address purely coherent transport on highly symmetric graphs (complete graph, complete bipartite graph, and star graph), and show analytically that it is possible to enhance the transport efficiency without environmental noise, i.e., using only a minimal perturbation of the graph. 
In a complete graph it is known that removing the edge between the vertex where the walker is initially localized and the trap leads to unit transport efficiency \cite{caruso2009highly,novo2015systematic}. The role of removing more than one randomly picked edges from a complete graph has been investigated in \cite{kurt2022quantum}, in both the noiseless and noisy case, showing that the maximum efficiency attainable is obtained when the network topology is modified by severing the edge between the source and the sink. On the other hand, appending a complex phase to an edge of the graph breaks the time-reversal symmetry of the unitary dynamics of a CTQW and results in a continuous-time chiral quantum walk. This can enable directional control, enhancement, and suppression of quantum transport \cite{zimboras2013quantum}. Indeed, complex-valued edge weights in a graph can completely suppress the flow of probability amplitude to specific vertices as discussed in \cite{sett2019zero,chaves2022and}. Other applications, instead, include quantum search \cite{wong2015quantum} and universal state transfer on graphs \cite{cameron2014universal}, where using complex Hermitian adjacency matrices make the CTQW chiral. Experimentally, continuous-time chiral quantum walks can be realized using nuclear magnetic resonance techniques in qubit systems \cite{PhysRevA.93.042302}.

In the present work, we show that adding an extra complex weight to one or two edges breaks the symmetries of the graph and generally enhances the transport efficiency \cite{fri21,fri22}. The exact nature of the perturbation depends on whether the initial state is localized or it is prepared in a superposition of two vertex states. {Our perturbation can be interpreted as a minimal disorder to coherently achieve the optimal transport efficiency. Therefore, in this sense our work is somehow linked to disorder-assisted quantum transport \cite{mohseni2013geometrical,novo2016disorder,zerah2020effects}, where disorder affecting on-site energies and/or couplings is intended as a tool for optimizing transport efficiency.} Our results indicate that breaking the symmetries of a graph may improve the transport efficiency. In particular, 
unit transport efficiency is obtained when the perturbation increases the dimension of
the relevant Krylov subspace. 

The paper is structured as follows: in Sec. \ref{sec:QWs} we introduce notation for quantum walk on graphs, and briefly review the concepts of dimensionality reduction and transport efficiency, which represent the relevant tools to analyze and assess transport properties of quantum walks on graphs.
In Sec. \ref{sec:Min_Per_App}, we introduce the minimal perturbation approach to optimize coherent transport and illustrate the procedure used to achieve unit transport efficiency on  graphs. In Sec. \ref{sec:Per_Qua_Tra}, we show results for different classes of graphs, both for localized initial states and for initial superposition states. In Sec. \ref{s:null}, we address the complementary problem of finding the conditions to achieve zero  transport efficiency, i.e., to avoid any loss phenomena. Finally, in Sec.  \ref{conclusion} we draw our conclusions. 

\section{Coherent transport on graphs: transport efficiency and dimensionality reduction}
\label{sec:QWs}
A graph is an ordered pair $G = \left( V, E \right)$ where $V$ is the set of vertices and $E$ the set of edges. It is mathematically represented by the Laplacian matrix $\mathcal{L}=\mathcal{D}-\mathcal{A}$, where $\mathcal{D}$ is the diagonal degree matrix of elements $\mathcal{D}_{jj} = \operatorname{deg}(j)$, degree of the vertex $j$, and $\mathcal{A}$ is the adjacency matrix, whose element $\mathcal{A}_{jk}$ is 1 if and only if the two vertices $j$, $k$ are connected, 0 otherwise. Hence, the generic element of the Laplacian matirx is
\begin{equation}
\mathcal{L}_{jk} = \begin{cases}
\operatorname{deg}(j) & \text{if $j=k$,} \\
-1 & \text{if  $j \neq k$ and $(j, k) \in E$,}\\
0 & \text{otherwise.} 
\end{cases}
\end{equation}

The vertices of the graphs represent the possible positions of the walker, whereas the edges correspond to the possible transitions among the vertices. The order of the graph is the number of vertices $\vert V \vert = N$. The states $\left\{\ket{j}\right\}_{j=1,...,N}$, where $j \in V$ is a vertex of $G$, form an orthonormal basis and span the $N$-dimensional Hilbert space of the quantum walker. 
In principle, a CTQW is generated by any Hamiltonian $\mathcal{H}$ (or, generally, any Hermitian operator) that respects the topology of the graph \cite{fri21,fri22}. The generic state $ |\psi(t)\rangle$ of the walker satisfies the Schr\"{o}dinger equation (we set $\hbar=1$ throughout the work)
\begin{equation}
i\frac{d}{dt}\langle a \vert \psi\left(t\right) \rangle = \sum_{b \in V} \langle a\vert \mathcal{H} \vert b\rangle\langle b\vert \psi\left(t\right)\rangle\,,
\end{equation}
where $\langle a \vert \psi\left(t\right) \rangle$ is the wavefunction written in the discrete space, i.e., the probability amplitude of finding the walker at the vertex $a$. According to this choice, energy and time are dimensionless in the following. In particular, the prototype of CTQW is defined by taking the Laplacian matrix as Hamiltonian of the system, $\mathcal{H} = \mathcal{L}$. 
Unlike the classical random walk where the state is described in terms of probability distribution of the walker, in a CTQW the state is written in terms of the (complex) probability amplitudes $q_{a}(t)=\langle a | \psi(t) \rangle$. Given the initial state $\ket{\psi_0}$, the state at a later time $t>0$ is
\begin{equation}
\label{wav_tim_t}
\ket{\psi(t)}=\mathcal{U}(t)\ket{\psi_0}= e^{-i \mathcal{H} t}\ket{\psi_0}\,,
\end{equation}
where $\mathcal{U}(t)$ is the unitary time-evolution operator. The evolution is unitary as long as the Hamiltonian of the system is Hermitian, and therefore the process preserves the norm of the state, $\sum_{a \in V} \left|\langle a |\psi\left(t\right)\rangle\right|^{2} = 1\,\forall t$.

Sometimes, to phenomenologically model certain processes, such as gain or loss processes, one can employ a non-Hermitian effective Hamiltonian. As a result, the time evolution is nonunitary and does not preserve the norm of the state.

In the present work, we model quantum transport as the CTQW of an excitation over a graph, where a single vertex $w$ absorbs the component $\bra{w} \psi(t)\rangle$ continuously in time with a rate $\kappa \in \mathbb{R}^+$. The vertex $w$ plays the role of a sink and we will refer to it as the trap (vertex). This mimics what happens, e.g., in photosynthetic systems, where an initial excitation is created by absorbing a solar photon and it is transferred along the light-harvesting complex to a reaction center where it gets absorbed and converted into chemical energy \cite{engel2007evidence,collini2010coherently}. In such a scenario, the total probability of finding the initial excitation within the network is not conserved. To achieve such a nonunitary dynamics we consider the non-Hermitian effective Hamiltonian
\begin{equation}
\mathcal{H}_{t}=\mathcal{L}-i\kappa \vert w \rangle \langle w \vert\,,
\label{eq:transport_Ham}
\end{equation}
where the anti-Hermitian term $-i\kappa \vert w \rangle \langle w \vert$ is responsible for the loss processes at $w$. We will refer to $\mathcal{H}_{t}$ as the {\em transport Hamiltonian}.

The transport efficiency is a commonly used figure of merit to assess the transport properties of the system. Intimately related to the total loss of  probability, it is defined as the integrated probability of trapping at the vertex $w$ in the limit of infinite time,
\begin{equation}
\label{eq:tr_eff_def}
\eta = 2 \kappa \int_{0}^{+\infty} \bra{w} \rho(t) \ket{w} dt = 1- \Tr \Big[ \lim_{t \rightarrow + \infty} \rho(t) \Big]\,
\end{equation}
where $ 2 \kappa \bra{w} \rho(t) \ket{w} dt $ is the probability that the walker is successfully absorbed at the trap within the time interval $ [ t, t + dt ] $ and $ \rho(t) = | \psi (t) \rangle \langle \psi(t) | $ is the density matrix of the walker. The surviving total probability of finding the walker within the graph at time $t$ is $ \Tr \left[ \rho(t) \right] = \langle \psi (t) | \psi(t) \rangle \leq 1 $ because of the loss processes at the trap vertex. According to this and because of the definition of transport efficiency, we can assess the transport efficiency also as the complement to 1 of the probability of surviving within the graph, which is the complementary event. Note that the value of $\eta$ does not actually depend on $\kappa$, since it is computed in the limit of long times.  

In a CTQW problem the quantity of interest is often the probability amplitude at a particular vertex of the graph. This is also our case, as the transport efficiency \eqref{eq:tr_eff_def} requires the probability amplitude $\braket{w}{\psi(t)}$. Pedantically solving the original problem, i.e., the eigen-problem for the Hamiltonian, to compute that amplitude can be a hard task, since the size of the problem grows linearly with the number of vertices (order), $N$, of the graph.
Nevertheless, as often happens in physics, we can exploit the symmetries of the problem, i.e., of the graph, to substantially simplify the original problem reducing its effective dimension. In fact, the dynamics relevant to our problem is entirely contained in a subspace of the full $N$-dimensional Hilbert space $\mathscr{H}$ and the dimension of such a subspace does not depend on $N$. 

This idea is formalized using the dimensionality reduction method \cite{novo2015systematic}, which we briefly review in the following. Let us consider the Taylor expansion of the time-evolution operator, the probability amplitude at $w$ can be written as
\begin{align}
\bra{w}e^{-i \mathcal{H}t}\ket{\psi_{0}}&=\sum_{k=0}^\infty \frac{(-it)^k}{k!}\bra{w}\mathcal{H}^k\ket{\psi_0}\nonumber\\
&=\bra{w}e^{-i\mathcal{H}_\textup{red} t}\ket{{\psi_0}_\textup{red}}\,,
\end{align}
where  $ \mathcal{PHP}=\mathcal{H}_\textup{red} $ is a reduced Hamiltonian, and $\ket{{\psi_0}_\textup{red}}= \mathcal{P}\ket{\psi_0}$ a reduced initial state, $\mathcal{P}$ being
the projector onto the Krylov subspace, which itself is defined as
\begin{equation}
\mathcal{I}(\mathcal{H},\ket{w}) = \operatorname{span}(\lbrace \mathcal{H}^k \ket{w} \mid k \in \mathbb{N}_0\rbrace)\,.
\label{eq:Krylov_subspace}
\end{equation}
Clearly, $\dim \mathcal{I}( \mathcal{H},\ket{w}) \leq \dim \mathscr{H}=N$, as $\mathcal{I}( \mathcal{H},\ket{w}) \subseteq \mathscr{H}$. However, due to the symmetries of the system and thus of the Hamiltonian, the number of states $\mathcal{H}^k \ket{w}$ that are actually linearly independent can be much lower than $N$. An orthonormal basis, $\{\ket{e_1},\ldots,\ket{e_m}\}$, for the subspace $\mathcal{I}(\mathcal{H},\ket{w})$ can be constructed as follows. Starting from the state $\ket{e_1}=\ket{w}$, the successive basis states are obtained by applying $\mathcal{H}$ to the current basis state and orthonormalizing the result with respect to the previous basis states. The procedure stops when we find the minimum $m$ such that the state $\mathcal{H}\ket{e_{m}}$ is a linear combination of the previous states $\ket{e_1},\ldots,\ket{e_{m}}$. The resulting reduced Hamiltonian written in such a basis has a tridiagonal form. Indeed, according to this method, at each iteration the state $\mathcal{H}\ket{e_k}$ is a linear combination of $\ket{e_{k-1}}$, $\ket{e_{k}}$, and the new basis state $\ket{e_{k+1}}$ to be defined. The original problem is then mapped onto an equivalent one of lower dimension, $m \leq N$, that is governed by a tight-binding Hamiltonian of a line with $m$ sites.

It can be proved that the transport efficiency can be computed as
\begin{equation}
\eta = \sum_{k=1}^m \left\vert \braket{e_k}{\psi_0}\right\vert^2\,,
\label{eq:transport_eff_ideal}
\end{equation}
i.e., as the overlap of the initial state $\ket{\psi_0} $ with the subspace $\mathcal{I}(\mathcal{H},\ket{w})$ \cite{novo2015systematic}. Indeed, such a subspace is the subspace spanned by the eigenstates of the Hamiltonian having a nonzero overlap with the trap $\ket{w}$ required to compute $\eta$ \cite{caruso2009highly}.

{It is important to emphasize that the dimensionality reduction method leads to a Krylov subspace whose dimension may be lower than that of the subspace spanned by the states obtained by grouping together the identically evolving vertices \cite{wang2021role}, i.e., those identified from the symmetries of the graph. In some cases the two subspaces coincide, in others they do not. Indeed, the subspace obtained by means of the dimensionality reduction method provides the subspace of minimum dimension which is relevant for the given problem of interest, here computing the probability at the trap vertex. The other approach, instead, is problem-independent in the sense that it relies only upon the symmetries of the graph (actually, of the Hamiltonian) and thus it provides a subspace with an equal or larger dimension. Therefore, the resulting subspace in general cannot be used to assess the transport efficiency according to \eqref{eq:transport_eff_ideal}.}

\section{Minimal perturbation approach}
\label{sec:Min_Per_App}
For an initially localized state or a superposition of two vertex states evolving under the transport Hamiltonian \eqref{eq:transport_Ham} it is unlikely to achieve a high transport efficiency. In addition, the transport efficiency $\eta$ usually decreases with the order $N$ of a graph \cite{razzoli2021transport}. Our starting point is the observation that breaking the symmetries of the graph can actually improve the transport efficiency. The first evidence of this behavior may be found in \cite{novo2015systematic}, where it was shown that removing the edge between the initial vertex and the trap in a complete graph leads to the unit transport efficiency.

In turn, the main goal of the present work is to put forward a general scheme to engineer the Hamiltonian \eqref{eq:transport_Ham} and achieve unit or nearly unit transport efficiency with a minimal perturbation of the graph. In particular, we are interest in those situations where the initial state is a localized state, $\ket{\psi_0} = \ket{l} $, or a superposition of two vertex states, $\ket{\psi_0}=(\ket{l}+e^{i\gamma}\ket{k})/\sqrt{2}$ with $0 \leq \gamma < 2 \pi$, assuming that the initial state does not involve the trap, thus $l,k \neq w$. To this aim,  we perturb some edges of the graph by adding and extra weight to them in the form $\lambda e^{i\theta}$, where $\lambda > 0$ and $0 \leq \theta < 2 \pi$ \cite{fri21,fri22}. If we perturb the edge $(r,s)$, then the corresponding Hamiltonian matrix elements are $\mathcal{H}_{rs}=-1+\lambda e^{i\theta}$ and $\mathcal{H}_{sr} = \mathcal{H}_{rs}^\ast$. The non-Hermiticity of the Hamiltonian still arises only from the trap. Denoting by $E_p \subseteq E$ the subset of perturbed edges, the complete Hamiltonian is
\begin{align}
\mathcal{H}_{c} &=\mathcal{H}_{t} +\sum_{(r,s)\in E_p} \left(  \lambda e^{i\theta_{rs}} \vert r \rangle \langle s \vert + \lambda e^{-i\theta_{rs}} \vert s \rangle \langle r \vert \right)\,,
\label{eq:engineered_Ham}
\end{align}
Note that setting $(\lambda = 1, \theta_{rs}=0)$ means removing the corresponding edge $(r,s)$, since the matrix element of the complete Hamiltonian is $ \mathcal{H}_{rs}=-1+\lambda e^{i \theta_{rs}}=\mathcal{H}_{sr}^\ast$. This model allows a perturbative approach due to the presence of the parameter $\lambda$, as $\mathcal{H}_{c}$ recovers $\mathcal{H}_{t}$ \eqref{eq:transport_Ham} in the limit of $\lambda  \to 0$. Notice that just appending a simple phase factor on an edge, i.e., defining $\mathcal{H}_{rs}=- e^{i\theta}=\mathcal{H}_{sr}^\ast$ would be sufficient to achieve unit transport efficiency in the complete graph, but would fail for other graphs, e.g., the star graph. This is why we do not simply add a phase factor to an existing edge, but rather add a full extra complex weight to the edge.  
In the following, we are going to consider some relevant classes of graphs and seek for the minimal subsets $E_p$ which allow one to achieve unit transport efficiency. As we will see, this is possible for complete and bipartite complete graphs, whereas for the specific case of the star graph, the procedure is successful with some constraints.
\section{Optimizing transport efficiency by minimal perturbations}
\label{sec:Per_Qua_Tra}
In this section we apply the dimensionality reduction method and the minimal perturbation approach to the Hamiltonian \eqref{eq:engineered_Ham} and the initial states considered above (localized or superposition of two vertex states) to evaluate 
the transport efficiency $\eta(\lambda ,\theta)$ of some classes of graphs, and find the conditions for which it is equal to 1. We consider some paradigmatic graphs in terms of connectivity and symmetry, namely the complete graph, the complete bipartite graph, and the star graph (particular case of complete bipartite graph). For each graph, we assess the transport efficiency first for the (unperturbed) transport Hamiltonian \eqref{eq:transport_Ham} and then for the perturbed Hamiltonian \eqref{eq:engineered_Ham}, properly engineered according to the given initial state.

\subsection{Complete graph}
\begin{figure}[htb]
	\centering
	\includegraphics[width=0.9\linewidth]{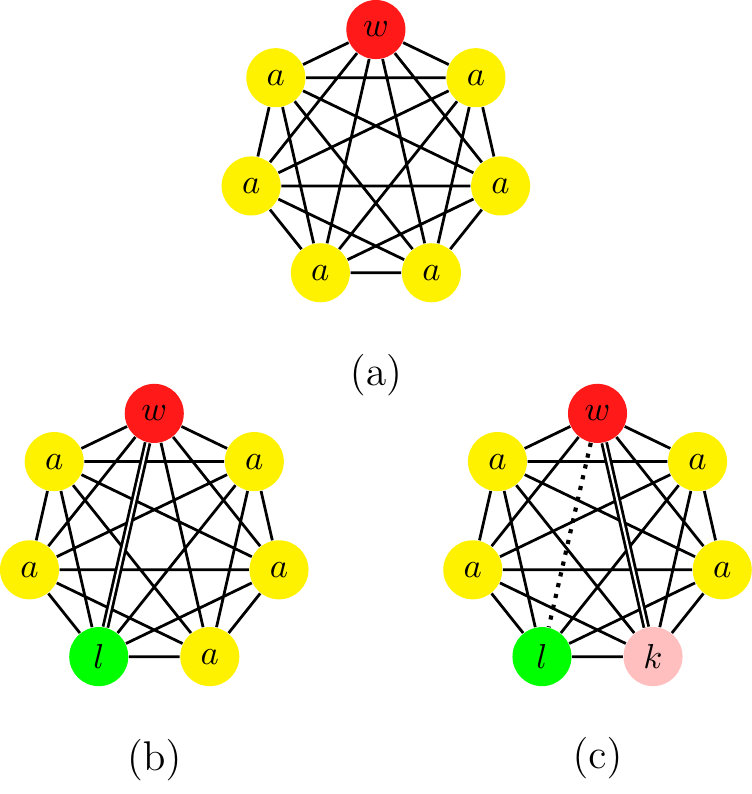}
	\caption{Complete graph of order $7$, $K_7$, for quantum transport. The trap vertex $w$ is colored in red. (a) Unperturbed graph. Engineered graphs to achieve the maximum transport efficiency  (b) for a localized initial state and Hamiltonian \eqref{eq:ls_cg} and (c) for an initial superposition of two vertices and Hamiltonian \eqref{eq:ss_cg}. The perturbed edges are denoted by doubled and dotted lines. Identically evolving vertices under the given Hamiltonian  are identically colored and labeled.}
	\label{fig:CG}
\end{figure}

In the complete graph of order $N$, $K_{N}$, each vertex is connected to all the others, thus it has degree $N-1$, and therefore the graph is regular, Fig.\ref{fig:CG}(a). The Laplacian matrix of $K_N$ is
\begin{equation}
\label{eq:Lapl_cg}
\mathcal{L}=N\mathbb{1}-\sum_{j,h}\vert j \rangle \langle h \vert\,,
\end{equation}
where $\mathbb{1}$ is the $ N \times N$ identity matrix.
The complete graph has one symmetry: the graph, as well as its Laplacian matrix, is invariant under the permutation of all its vertices. Making one vertex the trap breaks such a symmetry and the resulting transport Hamiltonian \eqref{eq:transport_Ham} is invariant under the permutation of all the vertices but the trap. 

The subspace \eqref{eq:Krylov_subspace} for the transport Hamiltonian is two-dimensional and is spanned by the basis states (see Appendix \ref{app:basis_cg})
\begin{align}
\ket{e_1} &= \ket{w}\,, \nonumber\\
\ket{e_2} &= \dfrac{1}{\sqrt{N-1}} \sum_{j \neq w} \ket{j}\,.
\label{eq:basis_cg}
\end{align}
Since the initial state does not involve the trap and all the remaining vertices in the complete graph are equivalent, according to \eqref{eq:transport_eff_ideal} the transport efficiency for any initial localized state $\ket{l \neq w}$ is
\begin{align}
\label{eq:eta_cg_l}
\eta = \left\vert \braket{e_2}{l}\right\vert^2 = \dfrac{1}{N-1}\,,
\end{align}
while for the superposition of two vertex states $(\ket{l} + e^{i\gamma}\ket{k})/\sqrt{2}$ is
\begin{align}
\label{eq:eta_cg_sup}
\eta = \frac{1}{2}\left( \left\vert \braket{e_2}{l}+e^{i\gamma}\braket{e_2}{k}\right\vert^2 \right) = \dfrac{1+ \cos\gamma}{N-1}\,.
\end{align}
The walker is completely absorbed at the trap ($\eta=1$) for $N=2$ if initially localized and for $N=3$ and $\gamma = 0$ for the initial superposition of states. In both cases, $\eta \sim 1/N$ for $N\gg 1$ and $\lim_{N \to +\infty} \eta = 0$.

\subsubsection{Localized Initial State}
To improve the transport efficiency for a localized initial state $\ket{l}$, we perturb the edge $(l,w)$ which connects the starting vertex and the trap vertex $w$, Fig.\ref{fig:CG}(b). The complete Hamiltonian \eqref{eq:engineered_Ham} is therefore
\begin{equation}
\label{eq:ls_cg}
\mathcal{H}_{c} = \mathcal{H}_{t} + \lambda e^{-i\theta}\vert l \rangle \langle w \vert + \lambda e^{i\theta}\vert w \rangle \langle l \vert.
\end{equation}
The perturbation breaks the symmetry of the complete graph with one trap. This is also reflected in the Krylov subspace $\mathcal{I}(\mathcal{H},\ket{w})$ (see Appendix \ref{app:basis_per_cg}) whose dimension changes from two, for the unperturbed transport Hamiltonian, to three and is spanned by the basis states
\begin{align}
\label{eq:basis_per_cg}
\ket{e_1} &= \ket{w} \nonumber\,,\\
\ket{e_2} &= \frac{ -\sum_{j \neq w,l} \ket{j} + \left( \lambda e^{-i\theta} - 1 \right) \ket{l} }{\sqrt{N-2 + \left| \lambda e^{-i\theta} -1 \right|^{2}}}\,, \nonumber\\
\ket{e_3} &= \frac{ \left( \lambda e^{i\theta}-1 \right) \sum_{j \neq w,l} \ket{j} + ( N-2) \ket{l} }{\sqrt{ ( N-2 ) (N-2 + \left| \lambda e^{-i\theta} -1 \right|^{2})}} \,.
\end{align}

Alternatively, as long as $\lambda \neq 0$ (the perturbation must exist), we can define a more suitable basis for $\mathcal{I}(\mathcal{H},\ket{w})$
\begin{align}
\label{eq:modified_basis_per_cg}
&\vert e_{1}' \rangle = \ket{w}\,, \nonumber\\
&\vert e_{2}' \rangle = \dfrac{1}{\sqrt{N-2}}\sum_{j \neq w,l} \ket{j}\,, \nonumber\\
&\vert e_{3}' \rangle = \ket{l}\,,
\end{align}
obtained by grouping together the vertices which evolve identically under the Hamiltonian considered and are identified by symmetry \cite{wang2021role}. The basis \eqref{eq:modified_basis_per_cg} does not depend on $\lambda$ or $\theta$ and it is possible to prove that any basis state of \eqref{eq:modified_basis_per_cg} can be written as a linear combination of the states \eqref{eq:basis_per_cg} and vice versa.

Therefore, the transport efficiency \eqref{eq:transport_eff_ideal} for the initial localized state $\ket{l}$ is
\begin{align}
\eta  &= \left| \langle e_2 \ket{l} \right|^2 + \left| \langle e_3 \ket{l} \right|^2
= \left| \langle e_3' \ket{l} \right|^2 = 1 \,,
\end{align}
independently of $\lambda $ or $\theta$. For the other vertices $\vert a \rangle $ with $a \neq w,l$, the transport efficiency is
\begin{align}
\eta &= \left| \langle e_2 \vert a \rangle \right|^2 + \left| \langle e_3 \vert a \rangle \right|^2 = \left| \langle e_2' \ket{l} \right|^2= \dfrac{1}{N-2}\,,
\end{align}
independently of $\lambda $ or $\theta$.

\subsubsection{Superposition State}
To improve the transport efficiency for an initial superposition of two vertex states, $( \ket{l} + e^{i\gamma}\ket{k})/\sqrt{2}$, we perturb the edges $(l,w)$ and $(k,w)$ which connect the two vertices of the initial state and the trap vertex $w$, Fig.\ref{fig:CG}(c). The complete Hamiltonian \eqref{eq:engineered_Ham} is therefore
\begin{align}
\label{eq:ss_cg}
\mathcal{H}_{c} = \mathcal{H}_{t} & + \lambda e^{-i\theta}\vert k \rangle \langle w \vert + \lambda e^{i\theta}\vert w \rangle \langle k \vert\\
& + \lambda \vert l \rangle \langle w \vert + \lambda \vert w \rangle \langle l \vert\, \nonumber.
\end{align}
We consider the phase factor only for one edge to have the minimum number of degrees of freedom.
The perturbation breaks the symmetry of the complete graph with one trap. This is also reflected in the Krylov subspace $\mathcal{I}(\mathcal{H},\ket{w})$ (see Appendix \ref{app:basis_per_cg}) whose dimension changes from two, for the unperturbed transport Hamiltonian, to three and is spanned by the basis states
\begin{align}
	\label{eq:basis_sup_per_cg}
	&\ket{e_1} = \ket{w}\,, \nonumber\\
	&\ket{e_2} = \dfrac{ -\sum_{j \neq w,l,k} \ket{j} + \left( \lambda  - 1 \right) \ket{l} + \left( \lambda e^{-i\theta} - 1 \right) \ket{k} }{\sqrt{N-2 + \left| \lambda e^{-i\theta} -1 \right|^{2} + \left| \lambda  - 1 \right|^2}}\,, \nonumber\\
	&\ket{e_3} = \dfrac{1}{\mathcal{N}} \left( \alpha_1 \sum_{j \neq w,l,k} \ket{j} + \alpha_2 \ket{l} + \alpha_3 \ket{k}  \right)\,, 
\end{align}
with 
\begin{align}
\label{eq:par_sup_per_cg}
\alpha_1 &= e^{-i\theta}+1-2\lambda \,,\nonumber\\\
\alpha_2 &= e^{-i\theta}+ \lambda \left(e^{i\theta} -1 \right)+2-N\,, \nonumber \\
\alpha_3 &=  e^{-i\theta} (2+\lambda-N)+1 - \lambda\,, \nonumber \\
\mathcal{N} &= \sqrt{(N-3)|\alpha_1|^2 + |\alpha_2|^2 + |\alpha_3|^2 } \,.
\end{align}

Similarly to what is done for an  initial localized state, as long as $\lambda \neq 0$, we can define a more suitable basis for $\mathcal{I}(\mathcal{H},\ket{w})$
\begin{align}
	\label{eq:modified_basis_sup_per_cg}
	\vert e_{1}' \rangle &= \ket{w}, \nonumber\\
	\vert e_{2}' \rangle &= \dfrac{-\sum_{j \neq w,l,k} \ket{j} + \frac{1}{2} \left[ \left(e^{i\theta}-1\right) \ket{l} + \left(e^{-i\theta}-1\right) \ket{k} \right]}{\sqrt{N-3 + 2\sin^2(\theta/2)}}, \nonumber\\
	\vert e_{3}' \rangle &= \dfrac{1}{\sqrt{2}} \left(\ket{l} + e^{-i\theta}\ket{k}\right)\,. 
\end{align}
The basis in \eqref{eq:modified_basis_sup_per_cg} does not depend on $\lambda$ but does depend on $\theta$. It is possible to prove that any state in \eqref{eq:modified_basis_sup_per_cg} may be written as a linear combination of the states in \eqref{eq:basis_sup_per_cg} and vice versa.
Using \eqref{eq:modified_basis_sup_per_cg}, the transport efficiency \eqref{eq:transport_eff_ideal} for the initial superposition of two vertex states may be written as
\begin{align}
\eta (\theta) &= \left| \langle e_2' \ket{l} \right|^2 + \left| \langle e_3' \ket{l} \right|^2 \nonumber\\
&= \cos^2[(\gamma+\theta)/2] + \frac{2\sin^2(\theta/2)\sin^2[(\gamma+\theta)/2]}{N-3 + 2\sin^2(\theta/2)}\,,
\end{align}
which is equal to 1 for
\begin{align}
\frac{\gamma + \theta}{2} = n\pi \quad (n \in \mathbb{N}_{0})\,,
\end{align}
i.e., for $\theta = 2 \pi - \gamma$ if we restrict the phases in $[0,2\pi[$ or for $\theta = - \gamma$ if we restrict the phases in $[-\pi,\pi[$.

\subsection{Complete bipartite graph}
\begin{figure}[htb]
	\centering
	\includegraphics[width=0.9\linewidth]{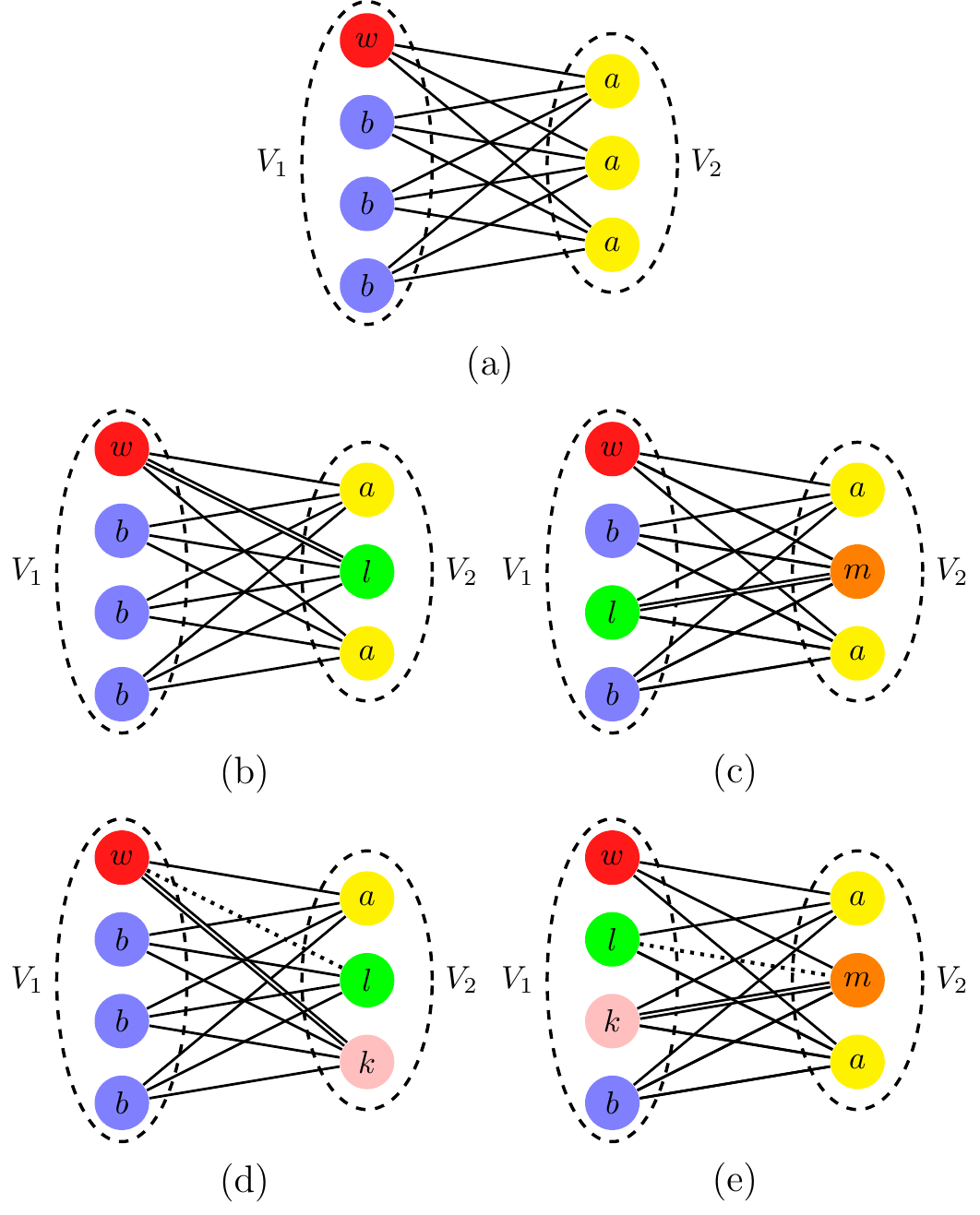}
	\caption{Complete bipartite graph $K_{4,3}$ of order $7$ for quantum transport. The trap vertex $w$ is colored in red. Vertices in $V_1$ and $V_2$ are respectively, second- and first-nearest neighbors of the trap $w$. (a) Unperturbed graph. Engineered graphs to achieve the maximum transport efficiency  for the initial state (b) localized at $l \in V_2$ and Hamiltonian \eqref{eq:ls_V2_cbg}; (c) localized at $l \in V_1$ and Hamiltonian \eqref{eq:ls_V1_cbg}; (d) superposition of two vertex states $l,k \in V_1$ and Hamiltonian \eqref{eq:sup_cbg_V1}; (e) superposition of two vertex states $l,k \in V_2$ and Hamiltonian \eqref{eq:sup_cbg_V2}. The perturbed edges are denoted by doubled and dotted lines. Identically evolving vertices under the given Hamiltonian  are identically colored and labeled.}
	\label{fig:CBG}
\end{figure}

The complete bipartite graph (CBG), $K_{N_1,N_2}$, is a highly symmetrical structure defined by two sets of vertices, $V_1$ and $V_2$, such that each vertex of $V_{1}$ is connected only to all the vertices of $V_{2}$ and vice versa, Fig.\ref{fig:CBG}(a). The number of vertices in $V_1$ and $V_2$ is respectively $N_{1}=\vert V_{1} \vert$ and $N_{2}= \vert V_{2} \vert$, so $N=N_{1}+N_{2}$ is the total number of vertices. Each vertex in $V_{1}$ has degree $N_{2}$, and each vertex in $V_{2}$ has degree $N_{1}$, thus the CBG is regular only if $N_{1} = N_{2}$.
The Laplacian matrix of $K_{N_{1},N_{2}}$ is
\begin{align}
\label{eq:Lapl_cbg}
\mathcal{L}=N_{2}\sum_{j \in V_{1}}\vert j \rangle \langle j \vert & + N_{1}\sum_{h \in V_{2}}\vert h \rangle \langle h \vert \notag \\ & - \sum_{j \in V_{1}} \sum_{h \in V_{2}} ( \vert j \rangle \langle h \vert + \vert h \rangle \langle j \vert)\,.
\end{align}
The CBG has two symmetries: the graph, as well as its Laplacian matrix, is invariant under the permutation (1) of all the vertices in $V_1$ and (2) of all the vertices in $V_2$ . Making one vertex in $V_1$ the trap preserves the symmetry (2) but breaks the symmetry (1) and the resulting transport Hamiltonian \eqref{eq:transport_Ham} is invariant under the permutation of all the vertices in $V_1$ but the trap.

The subspace \eqref{eq:Krylov_subspace} for the transport Hamiltonian is three-dimensional and is spanned by the basis states (see Appendix \ref{app:basis_cbg})
\begin{align}
\label{eq:basis_cbg}
\ket{e_1} &=\ket{w}\,,\nonumber\\
\ket{e_2} &=\frac{1}{\sqrt{N_2}}\sum_{h \in V_2}\ket{h} \nonumber \,,\\
\ket{e_3}&=\frac{1}{\sqrt{N_1-1}}\sum_{\substack{j \in V_1,\\j\neq w}}\ket{j}\,.
\end{align}

According to \eqref{eq:transport_eff_ideal}, the transport efficiency for an initial localized state $\ket{l \neq w}$ is
\begin{equation}
\eta = \begin{dcases}
\frac{1}{N_1-1} & \text{if $l \in V_1$}\\
\frac{1}{N_2} & \text{if $l \in V_2$,}
\end{dcases}
\label{eq:eta_cbg_l}
\end{equation}
while for the superposition of two vertex states $(\ket{l} + e^{i\gamma}\ket{k})/\sqrt{2}$ is
\begin{equation}
\eta = \begin{dcases}
\frac{1+\cos\gamma}{N_1-1} & \text{if $l,k \in V_1$,}\\
\frac{1+\cos\gamma}{N_2} & \text{if $l.k \in V_2$,}\\
\frac{N_1+N_2-1}{2(N_1-1)N_2} & \text{if $l \in V_1$ and $k\in V_2$,}
\end{dcases}
\label{eq:eta_cbg_s}
\end{equation}
provided that $l,k \neq w$ \cite{razzoli2021transport}.
There is a finite number of cases for which the walker is completely absorbed at the trap ($\eta=1$). Regarding an initial localized state \eqref{eq:eta_cbg_l}, this occurs for $N_1=2$ if $l\in V_1$ and for $N_2=1$ (star graph) if  $l\in V_2$. Regarding an initial superposition of two vertex states \eqref{eq:eta_cbg_s}, this occurs for $N_1=3$ and $\gamma = 0$ if $l,k\in V_1$, for $N_2=2$ and $\gamma = 0$ if $l,k\in V_2$, and for $N_1=2$ and $N_2=1$ if $l \in V_1$ and $k\in V_2$.


{Two remarks are here in order, with reference to Eqs. \eqref{eq:eta_cbg_l}--\eqref{eq:eta_cbg_s}. First, when the initial state (localized or superposition of two vertex states) involves only vertices from $V_1$ ($V_2$), then $\eta$ does not depend on $N_2$ ($N_1$). Second, we may obtain unit transport efficiency only for the sets $V_1$ and $V_2$ having minimum cardinality, i.e., if these sets contain only the trap and the vertices involved in the initial state (provided the proper phase in case of superposition). In detail, for a localized state $\ket{l}$, we have $\eta = 1$  iff $N_1=2$ and $V_1=\{w,l\}$, or $N_2=1$ and $V_2=\{l\}$ {(star graph with outer trap)}. Similarly, for the superposition $(\ket{l}+e^{i\gamma} \ket{k})/\sqrt{2}$, we obtain $\eta=1$ iff $N_1=3$, $V_1=\{w,l,k\}$, $\gamma = 0$, or $N_2=2$, $V_2=\{l,k\}$, $\gamma = 0$, or $N_1=2$, $N_2=1$, $V_1=\{w,l\}$ and $V_2=\{k\}$.
Therefore, if the CBG is not strongly unbalanced\footnote{A CBG is balanced if $N_1=N_2$. We say that a CBG is strongly unbalanced if $N_1 \ll N_2$ or $N_1 \gg N_2$.} and both $N_1$ and $N_2$ are sufficiently large, then neither the localized state, nor the superposition of two vertex states, can achieve a high transport efficiency. Actually, we have 
$\eta=O(1/N)$ for both the initial states considered ($N_1 \approx N_2 = O(N)$, since $N_1+N_2=N$).}

\subsubsection{Localized Initial State}
All vertices in $V_1 \setminus \{w\}$ are equivalent among them, as well as all vertices in $V_2$ are equivalent among them. Hence there are only two possible initial localized states to be discussed. We prove that we achieve $\eta = 1$ for the initial localized state $\ket{l}$ with $l \in V_2$ perturbing the edge $(l,w)$, as well as for the initial localized states $\ket{l}$ with $l \in V_1$ and $\ket{m}$ with $m \in V_2$ perturbing the edge $(l,m)$.
\paragraph{Edge $(l \in V_2,w)$}
To improve the transport efficiency for a localized initial state $\ket{l}$ with $l\in V_2$, first-nearest neighbor of the trap, we perturb the edge $(l,w)$ which connects the starting vertex and the trap vertex $w$, Fig.\ref{fig:CBG}(b). The complete Hamiltonian \eqref{eq:engineered_Ham} is therefore
\begin{equation}
\label{eq:ls_V2_cbg}
\mathcal{H}_{c} = \mathcal{H}_{t} + \lambda e^{-i\theta}\vert l \rangle \langle w \vert + \lambda e^{i\theta}\vert w \rangle \langle l \vert. 
\end{equation}
The perturbation breaks one symmetry of the CBG with one trap. This is also reflected in the Krylov subspace $\mathcal{I}(\mathcal{H},\ket{w})$ (see Appendix \ref{app:basis_per_cbg}) whose dimension changes from three, for the unperturbed transport Hamiltonian, to four and is spanned by the basis states
\begin{align}
\label{eq:basis_per_cbg}
&\ket{e_1}=\ket{w}\,,\nonumber \\
&\ket{e_2} =\frac{ \sum_{h \in V_2 \setminus \{l\}}\ket{h}  + \left( \lambda e^{-i\theta} - 1 \right) \ket{l} }{\sqrt{N_2 -1+ \left| \lambda e^{-i\theta} - 1 \right|^{2}}}\,, \nonumber\\
&\ket{e_3}=\frac{1}{\sqrt{N_1-1}}\sum_{j \in V_1 \setminus \{w\}}\ket{j}\,, \nonumber \\
&\ket{e_4}= \frac{ \left( \lambda e^{i\theta} - 1 \right) \sum_{h \in V_2 \setminus \{l\}}\ket{h}  + \left( N_2 -1 \right) \ket{l} }{\sqrt{(N_2 -1)(N_2 -1+ \left| \lambda e^{i\theta} - 1 \right|^{2})}} \,.
\end{align}
Again, as long as $\lambda \neq 0$ (the perturbation must exist), we can define a more suitable basis for $\mathcal{I}(\mathcal{H},\ket{w})$
\begin{align}
\label{eq:modified_basis_per_cbg}
&\ket{e_1'}=\ket{w}\,,\nonumber\\
&\ket{e_2'} =\frac{1}{\sqrt{N_2 -1}}\sum_{h \in V_2 \setminus \{l\}}\ket{h}\,, \nonumber \\
&\ket{e_3'}=\frac{1}{\sqrt{N_1-1}}\sum_{j \in V_1\setminus \{w\}}\ket{j}\,, \nonumber \\
&\ket{e_4'}=\ket{l}\,,
\end{align}
which does not depend on $\lambda$ or $\theta$.

Therefore, the transport efficiency \eqref{eq:transport_eff_ideal} for the initial localized state $\ket{l}$ is 
\begin{align}
\eta = \left| \langle e_2 \ket{l} \right|^2 + \left| \langle e_4 \ket{l} \right|^2 =  \left| \langle e_4' \ket{l} \right|^2 = 1,
\end{align}
independently of $\lambda $ or $\theta$. If $\ket{\psi_0} = \vert a \rangle $ with $a \in V_2 \setminus \{l\}$, then
\begin{align}
\eta = \left| \langle e_2 \vert a \rangle \right|^2 + \left| \langle e_4 \vert a \rangle \right|^2 =  \left| \langle e_2' \vert a \rangle \right|^2 = \dfrac{1}{N_2-1}\,.
\end{align}
If $ \vert \psi_{0}=\vert b \rangle$  with $b \in V_1 \setminus \{w\}$, then 
\begin{align}
\eta = \left| \langle e_3 \vert b \rangle \right|^2  =  \left| \langle e_3' \vert b \rangle \right|^2 = \dfrac{1}{N_1-1}\,,
\end{align}
the same result of the unperturbed case \eqref{eq:eta_cbg_l} and, indeed, the perturbation does not affect the vertices in $V_1$.

\paragraph{Edge $(l\in V_1,m \in V_2)$ } To improve the transport efficiency for a localized initial state $\ket{l}$ with $l\in V_1$, second-nearest neighbor of the trap, there is no need to create an edge connecting $l$ and the trap vertex $w$. It is sufficient to perturb the edge $(l,m)$ which connects the starting vertex and one of the vertices, say $m$, in $V_1$, Fig.\ref{fig:CBG}(c). The complete Hamiltonian \eqref{eq:engineered_Ham} is therefore
\begin{equation}
\label{eq:ls_V1_cbg}
\mathcal{H}_{c} = \mathcal{H}_{t} + \lambda e^{-i\theta}\ketbra{l}{m} + \lambda e^{i\theta}\ketbra{m}{l}. 
\end{equation}
The perturbation breaks both the symmetries of the CBG with one trap. This is also reflected in the Krylov subspace $\mathcal{I}(\mathcal{H},\ket{w})$ whose dimension changes from three, for the unperturbed transport Hamiltonian, to five and is spanned by the basis states
\begin{align}
	\label{eq:basis_per_cbg_2nn}
	&\ket{e_1}=\ket{w}\,,\nonumber \\
	&\ket{e_2}=\frac{1}{\sqrt{N_2}} \sum_{h \in V_2} \ket{h} \,,\nonumber \\
	&\ket{e_3}=\frac{-N_2 \sum_{j \in V_1 \setminus \{w,l\}} \ket{j} + \left( \lambda e^{i\theta} -N_2 \right) \ket{l}}{\sqrt{N_2^2 \left( N_1-2 \right) + \left| \lambda e^{i\theta} - N_2 \right|^2}} \,,\nonumber \\
    &\ket{e_4}=\frac{-\sum_{h \in V_2 \setminus \{m\} }\ket{h} + (N_2-1 ) \ket{m}}{\sqrt{N_2( N_2-1)}}\,, \nonumber \\
	&\ket{e_5}=\frac{\left( \lambda e^{-i\theta} - N_2 \right) \sum_{j \in V_1 \setminus \{w,l\}}\ket{j} + N_2\left( N_1 -2 \right)\ket{l} }{\sqrt{\left( N_1 - 2 \right)\left| \lambda e^{-i\theta} - N_2 \right|^2 + \left[ N_2 \left( N_1 - 2 \right) \right]^2}}\,.
\end{align}

Again, as long as $\lambda \neq 0$ (the perturbation must exist), we can define a more suitable basis for $\mathcal{I}(\mathcal{H},\ket{w})$
\begin{align}
	\label{eq:modified_basis_per_cbg_2nn}
	&\ket{e_1'}=\ket{w}\,, \nonumber\\
	&\ket{e_2'}=\frac{1}{\sqrt{N_2 -1}}\sum_{h \in V_2 \setminus \{m\}}\ket{h}\,, \nonumber \\
	&\ket{e_3'}=\ket{m}\,, \nonumber \\
	&\ket{e_4'}=\frac{1}{\sqrt{N_1-2}}\sum_{j \in V_1 \setminus \{w,l\} }\ket{j}\,, \nonumber \\
	&\ket{e_5'}=\ket{l}\,.
\end{align}
which does not depend on $\lambda$ or $\theta$.

Therefore, the transport efficiency \eqref{eq:transport_eff_ideal} for the initial localized state $\ket{l}$ is 
\begin{align}
	\eta = \left| \langle e_3 \ket{l} \right|^2 + \left| \langle e_5 \ket{l} \right|^2 = \left| \langle e_5' \ket{l} \right|^2 =1,
\end{align}
independently of $\lambda $ or $\theta$. If $\ket{\psi_0}=\ket{a}$ with $a \in V_2 \setminus \{m\}$, then
\begin{align}
	\eta = \left| \langle e_2 \vert a \rangle \right|^2 + \left| \langle e_4 \vert a \rangle \right|^2 = \left| \langle e_2' \vert a \rangle \right|^2 = \dfrac{1}{N_2-1}\,.
\end{align}
If $\ket{\psi_0}=\ket{b}$ with $b \in V_1 \setminus \{w,l\}$, then
\begin{align}
	\eta = \left| \langle e_3 \vert b \rangle \right|^2 + \left| \langle e_5 \vert b \rangle \right|^2 = \left| \langle e_4' \vert b \rangle \right|^2 = \dfrac{1}{N_1-2}\,.
\end{align}
If $\ket{\psi_0}=\ket{m}$, then
\begin{align}
	\eta = \left| \langle e_2 \vert m \rangle \right|^2 + \left| \langle e_4 \vert m \rangle \right|^2 = \left| \langle e_3' \vert m \rangle \right|^2 =1\,.
\end{align}
The latter result shows that we can achieve the maximum transport efficiency for an initial state $\ket{m}$ localized in $V_2$ without perturbing the edge $(m,w)$ connecting it to the trap, as done before (see Fig.\ref{fig:CBG}(b)).
Perturbing only one edge $(l,m)$ between vertices in different sets in the CBG, $l\in V_1$ and $m \in V_2$, allows us to achieve $\eta =1$ for two initial localized states, $\ket{l}$ and $\ket{m}$.

\subsubsection{Superposition State}
The CBG is defined by two distinct sets of vertices and because of their symmetries there are only three possible cases for an initial superposition of two vertex states $\ket{\psi_{0}}=\left(\ket{l}+e^{i\gamma}\ket{k}\right)/\sqrt{2}$.

\paragraph{$l,k \in V_2$} We consider the superposition of two vertex states that are first-nearest neighbors of the trap, Fig.\ref{fig:CBG}(d).  We perturb the edges $(l,w)$ and $(k,w)$ defining the complete Hamiltonian \eqref{eq:engineered_Ham} as
\begin{align}
\label{eq:sup_cbg_V1}
\mathcal{H}_{c} &= \mathcal{H}_{t} + \lambda e^{-i\theta}\vert k \rangle \langle w \vert + \lambda e^{i\theta}\vert w \rangle \langle k \vert  \\
&+ \lambda \vert l \rangle \langle w \vert + \lambda \vert w \rangle \langle l \vert. \nonumber
\end{align}

The perturbation breaks one symmetry of the CBG with one trap. This is also reflected in the Krylov subspace $\mathcal{I}(\mathcal{H},\ket{w})$ (see Appendix \ref{app:basis_per_cbg}) whose dimension changes from three, for the unperturbed transport Hamiltonian, to four and is spanned by the basis states
\begin{align}
\label{eq:basis_sup_per_cbg}
&\ket{e_1}=\ket{w} \,,\nonumber\\
&\ket{e_2} =\frac{ -\sum_{h \in V_2 \setminus \{l,k\}}\ket{h} + \left( \lambda - 1 \right) \ket{l} + \left( \lambda e^{-i\theta} - 1 \right) \ket{k} }{\sqrt{N_2 -2 + \left| \lambda e^{-i\theta} - 1 \right|^{2} + |\lambda  - 1|^2}} \,,\nonumber\\
&\ket{e_3}=\frac{1}{\sqrt{N_1-1}}\sum_{j \in V_1\setminus \{ w\}}\ket{j}\,, \nonumber \\
&\ket{e_4}= \dfrac{1}{\mathcal{N}'} \left( \alpha_1' \sum_{h \in V_2 \setminus \{l,k\}}^{N_{2}} \ket{j} + \alpha_2' \ket{l} + \alpha_3' \ket{k}  \right)\,,
\end{align}
with
\begin{align}
\label{eq:par_sup_per_cbg}
\alpha_1' &= e^{-i\theta} -1 -2\lambda \,, \nonumber \\
\alpha_2' &= e^{-i\theta}+ \lambda \left(e^{i\theta} -1 \right)+1-N_2\,, \nonumber \\
\alpha_3' &=e^{-i\theta}(1+\lambda-N_2)+ 1-\lambda\,, \nonumber\\
\mathcal{N}' &= \sqrt{(N_{2}-2)|\alpha_1'|^2 + |\alpha_2'|^2 + |\alpha_3'|^2 }\,.
\end{align}

Reasoning by symmetry, as long as $\lambda \neq 0$ (the perturbation must exist), we can define a more suitable basis for $\mathcal{I}(\mathcal{H},\ket{w})$
\begin{align}
\label{eq:modified_basis_sup_per_cbg}
&\ket{e_1'}=\ket{w}\,,\nonumber\\
&\ket{e_2'} =\dfrac{-\sum_{h\in V_2 \setminus \{l,k\}} \ket{h} + \frac{1}{2}\left[ (e^{i\theta}-1) \ket{l} + (e^{-i\theta}-1) \ket{k}\right]}{\sqrt{N_2-2 +2 \sin^2(\theta/2) }}\,, \nonumber \\
&\ket{e_3'}=\frac{1}{\sqrt{N_1-1}}\sum_{j \in V_1 \setminus \{w\}}\ket{j}\,, \nonumber \\
&\ket{e_4'}=\dfrac{1}{\sqrt{2}} \left( \ket{l} + e^{-i\theta}\ket{k} \right)\,,
\end{align}
which does not depend on $\lambda$ but does depend on $\theta$.

Therefore, according to the basis \eqref{eq:modified_basis_sup_per_cbg}, the transport efficiency \eqref{eq:transport_eff_ideal} for the initial superposition of two vertex states is
\begin{align}
\eta (\theta) &= \left| \langle e_2' \ket{l} \right|^2 + \left| \langle e_4' \ket{l} \right|^2 \nonumber\\
&= \cos^2[(\gamma+\theta)/2] + \frac{2\sin^2(\theta/2)\sin^2[(\gamma+\theta)/2]}{N_2-2 + 2\sin^2(\theta/2)}\,,
\end{align}
and equal to 1 (optimal) for
\begin{align}
\frac{\gamma + \theta}{2} = n\pi \quad (n \in \mathbb{N}_{0})\,,
\end{align}
i.e., for $\theta = 2 \pi - \gamma$ if we restrict the phases in $[0,2\pi[$ or for $\theta = - \gamma$ if we restrict the phases in $[-\pi,\pi[$.

\paragraph{$l,k \in V_1$}  We consider the superposition of two vertex states that are second-nearest neighbors of the trap, Fig.\ref{fig:CBG}(e). We perturb the edges $(l,m)$ and $(k,m)$ with $m \in V_2$ defining the complete Hamiltonian \eqref{eq:engineered_Ham} as
\begin{equation}
\label{eq:sup_cbg_V2}
\mathcal{H}_{c} = \mathcal{H}_{t} + \lambda e^{-i\theta}\ketbra{k}{m} + \lambda e^{i\theta}\ketbra{m}{k} + \lambda \ketbra{l}{m} + \lambda \ketbra{m}{l}\,. 
\end{equation}

The perturbation breaks both the symmetries of the CBG with one trap. This is also reflected in the Krylov subspace $\mathcal{I}(\mathcal{H},\ket{w})$ (see Appendix \ref{app:basis_per_cbg}) whose dimension changes from three, for the unperturbed transport Hamiltonian, to five and is spanned by the basis states
\begin{align}
\label{eq:basis_sup2_per_cbg}
&\ket{e_1}=\ket{w}\,,\nonumber \\
&\ket{e_2}=\frac{1}{\sqrt{N_2}}\sum_{\substack{h \in V_2}}\ket{h} \,, \nonumber \\
&\ket{e_3}=\frac{ -N_{2}\sum_{j \in V_1 \setminus \{w,l,k\}}\ket{j} + \left( \lambda - N_{2} \right) \ket{l} }{\sqrt{N_2^2\left(N_1 -3\right) + \left| \lambda e^{-i\theta} - N_{2} \right|^{2} + |\lambda  - N_{2}|^2}} \nonumber \\
& + \dfrac{\left( \lambda e^{-i\theta} - N_{2} \right) \ket{k}}{\sqrt{N_2^2\left(N_1 -3\right) + \left| \lambda e^{-i\theta} - N_{2} \right|^{2} + |\lambda  - N_{2}|^2}}\,, \nonumber \\
&\ket{e_4}=\dfrac{-\sum_{h \in V_2 \setminus \{ m\}}\ket{h} + (N_2 - 1) \ket{m}}{\sqrt{N_2(N_2-1)}}\,, \nonumber \\
&\ket{e_5}=\dfrac{1}{\mathcal{N}''} \left( \alpha_1'' \sum_{j \in V_1 \setminus \{ w,l,k\}} \ket{j} + \alpha_2'' \ket{l} +  \alpha_3'' \ket{k}  \right)\,,
\end{align}
with 
\begin{align}
\label{eq:par_sup2_per_cbg}
\alpha_1'' &= N_2(1+e^{-i\theta}) - 2\lambda\,, \nonumber\\
\alpha_2'' &= N_2\left[ \lambda (e^{i\theta}-1)+ N_2(e^{-i\theta}+2-N_1)\right]\,, \nonumber \\
\alpha_3'' &= N_2\left\lbrace N_2-\lambda+ e^{-i\theta}\left[\lambda-N_2(N_1-2)\right]\right\rbrace\,, \nonumber \\
\mathcal{N}'' &= \sqrt{(N_{1}-3)|\alpha_1''|^2 + | \alpha_2''|^2 + |\alpha_3''|^2 }\,.
\end{align}

Reasoning by symmetry, as long as $\lambda \neq 0$ (the perturbation must exist), we can define a more suitable basis for $\mathcal{I}(\mathcal{H},\ket{w})$
\begin{align}
\label{eq:modified_basis_sup2_per_cbg}
\ket{e_1'} &=\ket{w}\,,\nonumber\\
\ket{e_2'} &=\frac{1}{\sqrt{N_2-1}}\sum_{h \in V_2 \setminus \{m\}}\ket{h} \,,\nonumber \\
\ket{e_3'} &=\dfrac{-\sum_{j\in V_1 \setminus \{w,l,k\}} \ket{j} + \frac{1}{2}\left[ (e^{i\theta}-1) \ket{l} + (e^{-i\theta}-1) \ket{k}\right] }{\sqrt{N_1-3 +2 \sin^2(\theta/2)}} \,,\nonumber \\
\ket{e_4'} &=\ket{m}\,, \nonumber \\
\ket{e_5'} &=\dfrac{1}{\sqrt{2}} \left( \ket{l} + e^{-i\theta}\ket{k} \right)\,,
\end{align}
which does not depend on $\lambda$ but does depend on $\theta$.

Therefore, according to the basis \eqref{eq:modified_basis_sup2_per_cbg}, the transport efficiency \eqref{eq:transport_eff_ideal} for the initial superposition of two vertex states is
\begin{align}
\eta (\theta) &= \left| \langle e_3' \ket{l} \right|^2 + \left| \langle e_5' \ket{l} \right|^2 \nonumber\\
&= \cos^2[(\gamma+\theta)/2] + \frac{2\sin^2(\theta/2)\sin^2[(\gamma+\theta)/2]}{N_1-3 + 2\sin^2(\theta/2)}\,,
\end{align}
and equal to 1 (optimal) for
\begin{align}
\frac{\gamma + \theta}{2} = n\pi \quad (n \in \mathbb{N}_{0})\,,
\end{align}
i.e., for $\theta = 2 \pi - \gamma$ if we restrict the phases in $[0,2\pi[$ or for $\theta = - \gamma$ if we restrict the phases in $[-\pi,\pi[$.

\paragraph{$l\in V_1\,,k \in V_2$} In the last case, we consider the superposition of a first- and a second-nearest neighbor of the trap, perturbing the graph as in Fig.\ref{fig:CBG}(c). The complete Hamiltonian is \eqref{eq:ls_V1_cbg} and the Krylov subspace is spanned by the states \eqref{eq:modified_basis_per_cbg_2nn} (where the vertex $m$ is here the vertex $k$). Therefore, the transport efficiency \eqref{eq:transport_eff_ideal} for the initial superposition of two vertex states is
\begin{equation}
	\eta = \frac{1}{2} \left( \left| \bra{e_5'}\ket{l} \right|^2 + \left|e^{i\gamma}\braket{e_3'}{k} \right|^2 \right) = 1\,,
\end{equation}
independently of the perturbation, $\lambda $ and $\theta$, and of the phase $\gamma$ of the superposition.

\subsection{Star graph: Central trap}
\begin{figure}[htb]
	\centering
	\includegraphics[width=0.9\linewidth]{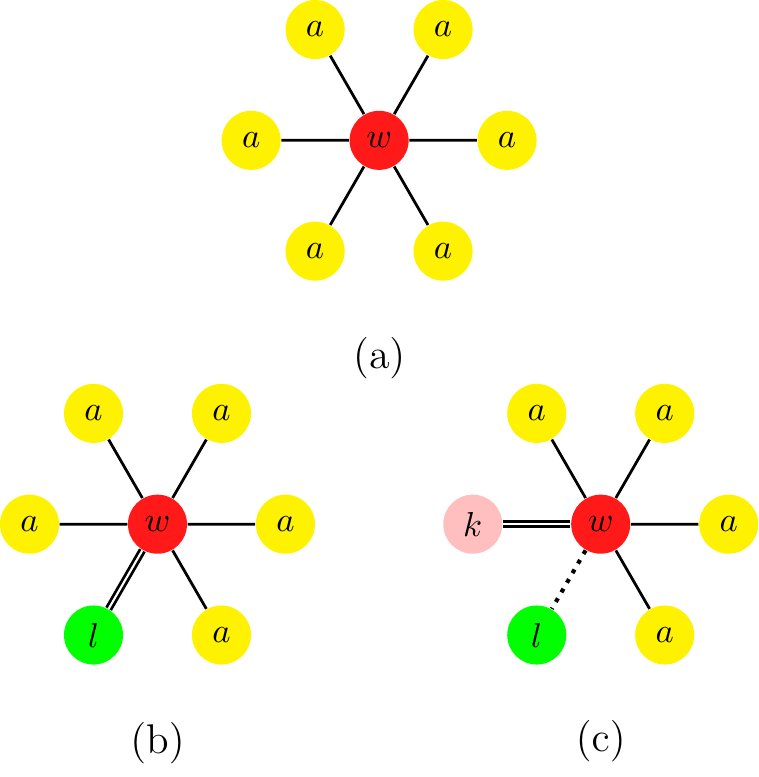}
	\caption{Star $S_7$ of order $7$ for quantum transport. The trap vertex $w$ is colored in red and is the central vertex (fully connected). (a) Unperturbed graph. Engineered graphs to achieve the maximum transport efficiency  (b) for a localized initial state and Hamiltonian \eqref{eq:ls_sg_ham} and (c) for an initial superposition of two vertices and Hamiltonian \eqref{eq:ss_sg_ham}. The perturbed edges are denoted by doubled and dotted lines. Identically evolving vertices under the given Hamiltonian  are identically colored and labeled.}
	\label{fig:SG}
\end{figure}

The star graph of order $N$, $S_N$, is a particular case of CBG with $V_1$ or $V_2$ having only one vertex. We assume  $N_1=1$, so $N_2=N-1$, and the central vertex as the trap; Fig.\ref{fig:SG}(a).
The Laplacian matrix of $S_{N}$ is
\begin{equation}
\label{eq:Lapl_sg_central}
\mathcal{L}=(N-1)\vert w \rangle \langle w \vert+\sum_{j \neq w}\vert j \rangle \langle j \vert - \sum_{j \neq w}( \vert j \rangle \langle w \vert + \vert w \rangle \langle j \vert)\,.
\end{equation}
The star graph has one symmetry: the graph, as well as its Laplacian matrix, is invariant under the permutation of all its outer vertices. Making the central vertex the trap preserves such a symmetry and the resulting transport Hamiltonian \eqref{eq:transport_Ham} is invariant under the permutation of all the outer vertices. 
Since the trap is a fully connected vertex, the subspace \eqref{eq:Krylov_subspace} for the transport Hamiltonian \eqref{eq:transport_Ham} is the same as that of the complete graph and is spanned by the states \eqref{eq:basis_cg} \cite{razzoli2022universality}. Therefore, the transport efficiency for an initial localized state is \eqref{eq:eta_cg_l} and that for an initial superposition of two vertex states is \eqref{eq:eta_cg_sup}.

\subsubsection{Localized Initial State}
To improve the transport efficiency for a localized initial state $\ket{l}$, we perturb the edge $(l,w)$ which connects the starting vertex and the trap vertex $w$, Fig.\ref{fig:SG}(b). The complete Hamiltonian \eqref{eq:engineered_Ham} is therefore
\begin{align}
\label{eq:ls_sg_ham}
\mathcal{H}_{c} &= \mathcal{H}_{t} + \lambda e^{-i\theta}\ket{l}\bra{w} + \lambda e^{i\theta}\ket{w}\bra{l}. 
\end{align}
The perturbation breaks the symmetry of the star graph with one trap. This is also reflected in the Krylov subspace $\mathcal{I}(\mathcal{H},\ket{w})$ (see Appendix \ref{app:basis_per_sg_cen_tra}), which, despite being two-dimensional as that for the unperturbed transport Hamiltonian, is spanned by different  basis states
\begin{align}
\label{eq:basis_sg_l}
\ket{e_1} &= \ket{w}\,,\nonumber\\
\ket{e_2} &= \dfrac{-\sum_{j \neq w,l} \ket{j} + \left( \lambda e^{-i\theta} - 1 \right) \ket{l}}{\sqrt{N-2 + \left| \lambda e^{-i\theta} -1 \right|^{2}}} \,.
\end{align}
The perturbation alters the Krylov subspace relevant to the computation of the transport efficiency since the two bases for the unperturbed and perturbed system, \eqref{eq:basis_cg} and \eqref{eq:basis_sg_l}, respectively, span different two-dimensional subspaces.

Therefore, the transport efficiency \eqref{eq:transport_eff_ideal} for the initial localized state $\ket{l}$ is
\begin{align}
\eta(\lambda,\theta) &= \left\vert \braket{e_2}{l}\right\vert^2 = \dfrac{\left| \lambda e^{-i\theta} - 1 \right|^2}{N-2 + \left| \lambda e^{-i\theta} -1 \right|^{2}}\nonumber\\
&=1-\frac{N-2}{\lambda^2-2\lambda \cos(\theta)+N-1}\,.
\label{eq:eta_sg_l}
\end{align}
The star graph $S_2$ ($N=2$, i.e., two vertices connected by an edge) provides $\eta =1$ independently of the perturbation. For $N>2$, instead, the transport efficiency does depend on the perturbation, both $\lambda$ and $\theta$, and its maximum is
\begin{equation}
    \eta_{max} = \eta(\lambda,\pi) = \frac{(\lambda + 1)^2}{\lambda^2+2\lambda+N-1}
\end{equation}
and, as evident also from \eqref{eq:eta_sg_l}, approaches the optimal value 1 in the limit of an infinitely strong perturbation
\begin{equation}
\lim_{\lambda \to +\infty} \eta \left( \lambda ,\theta \right) = 1\,.
\end{equation}
Therefore, unlike the other graphs, a small perturbation of the edge connecting the starting vertex to the trap does not lead to the optimal transport efficiency.

\subsubsection{Superposition State}
\label{subsec:SG_central_ss}
To improve the transport efficiency for an initial superposition of two vertex states, $( \ket{l} + e^{i\gamma}\ket{k})/\sqrt{2}$, we perturb the edges $(l,w)$ and $(k,w)$ which connect the two vertices of the initial state and the trap vertex $w$, Fig.\ref{fig:SG}(c). The complete Hamiltonian \eqref{eq:engineered_Ham} is therefore
\begin{align}
\label{eq:ss_sg_ham}
\mathcal{H}_{c} =& \mathcal{H}_{t} + \lambda e^{-i\theta}\ket{k}\bra{w} + \lambda e^{i\theta}\ket{w}\bra{k}\nonumber \\
&+\lambda \ket{l}\bra{w} + \lambda \ket{w}\bra{l}\,.
\end{align}
We consider the phase factor only for one edge to have the minimum number of degrees of freedom.

Similarly to the case of the localized state, the resulting Krylov subspace is two-dimensional and is spanned by the states (see Appendix \ref{app:basis_per_sg_cen_tra})
\begin{align}
\label{eq:basis_sup_sg_central}
\ket{e_1} &= \ket{w}\,,\nonumber\\
\ket{e_2} &= \dfrac{-\sum_{j \neq w,l,k} \ket{j} + \left( \lambda  -1 \right) \ket{l} +\left( \lambda e^{-i\theta}-1 \right) \ket{k}}{\sqrt{N-3 +\left| \lambda e^{-i\theta} - 1 \right|^{2}+ (\lambda  - 1)^{2}}}\,.
\end{align}
Therefore, the transport efficiency \eqref{eq:transport_eff_ideal} for the initial superposition of two vertex states is 
\begin{align}
\eta(\lambda,\theta,\gamma) &= \left\vert \braket{e_2}{\psi_0}\right\vert^2\nonumber\\
&= \dfrac{\left|\left( \lambda - 1 \right)+ e^{i\gamma}\left( \lambda e^{-i\theta}  - 1 \right) \right|^2}{2[N-3 + (\lambda - 1 )^{2} +\left| \lambda e^{-i\theta} - 1 \right|^{2}]}.
\end{align}
It depends not only on the perturbation, $\lambda$ and $\theta$, but also on the specific initial superposition, $\gamma$. 
{We observe that $\eta(\lambda,\theta=0, \gamma=0) =1$ for $N=3$, otherwise the transport efficiency is never optimal, $\eta <1$.
\begin{proof}
The transport efficiency can be rewritten as
\begin{equation}
    \eta(\lambda,\theta, \gamma) = \frac{\vert \alpha_l+ e^{i\gamma}\alpha_k \vert^2}{2(N-3)+2 (\vert \alpha_l \vert^2 +\vert \alpha_k \vert^2)}\,,
\end{equation}
where $\alpha_l = \lambda -1$ and $\alpha_k = \lambda e^{-i\theta} -1$. Now, we set $N=3$ and we study when $\eta =1$. Following the proof of the triangle equality we have
\begin{align}
\vert \alpha_l +e^{i\gamma}\alpha_k \vert^2 &\leq \vert \alpha_l \vert^2 +\vert \alpha_k \vert^2 + 2 \vert \alpha_l \vert \vert \alpha_k \vert\nonumber\\
&\leq 2 (\vert \alpha_l \vert^2 +\vert \alpha_k \vert^2)\,,
\end{align}
where the last inequality follows from 
$(\vert \alpha_l \vert -\vert \alpha_k \vert)^2 \geq 0$.
If $\vert \alpha_l \vert = \vert \alpha_k \vert =:\vert \alpha \vert$, then the inequality reduces to
\begin{align}
\vert \alpha \vert^2 \cos^2(\gamma/2) &\leq \vert \alpha \vert^2\,.
\end{align}
The equality holds for $\alpha = 0$ or, otherwise, for $\gamma = 0$.
In detail, recalling their definition, $\vert \alpha_l \vert = \vert \alpha_k \vert$ if $\theta = 0$. This is the first necessary condition to achieve $\eta=1$. The solution $\alpha = 0$ is not acceptable as it implies $\lambda = 1$, i.e., removing both the edges $(w,k)$ and $(w,l)$,\footnote{The matrix element of the complete Hamiltonian is $\bra{w}\mathcal{H}\ket{k}=\bra{k}\mathcal{H}\ket{w}=-1+\lambda$. Similarly, for $\bra{w}\mathcal{H}\ket{l}=\bra{l}\mathcal{H}\ket{w}$ because under the condition $\theta =0$.} i.e., all the edges of $S_3$. If $\lambda \neq 1$, the other solution is for $\gamma = 0$,  from $\cos^2(\gamma/2)=1$.
In conclusion, if $N=3$, then $\eta(\lambda,\theta=0, \gamma=0) =1$. Otherwise and for $N>3$, the transport efficiency is never optimal, $\eta <1$.
\end{proof}}
As well as in the case of the initial localized state, the optimal transport efficiency $\eta = 1$ is attained independently of $N$ in the limit of $\lambda \to +\infty$.


The star graph is a case of interest because, unlike all the previous examples investigated, the minimal perturbation is not sufficient to achieve the optimal transport efficiency. In all the previous cases, the perturbation affects the Krylov subspace required to assess the transport efficiency and increases the dimension of the subspace in such a way that the initial state considered is entirely contained in it. As a result, the overlap of the initial state and the subspace is equal to 1, thus $\eta =1$. Instead, in the star graph the perturbation changes the Krylov subspace but preserves its dimension and the initial state considered is never entirely contained in it.

\subsection{Star graph: Outer trap}
\begin{figure}[htb]
	\centering
	\includegraphics[width=0.9\linewidth]{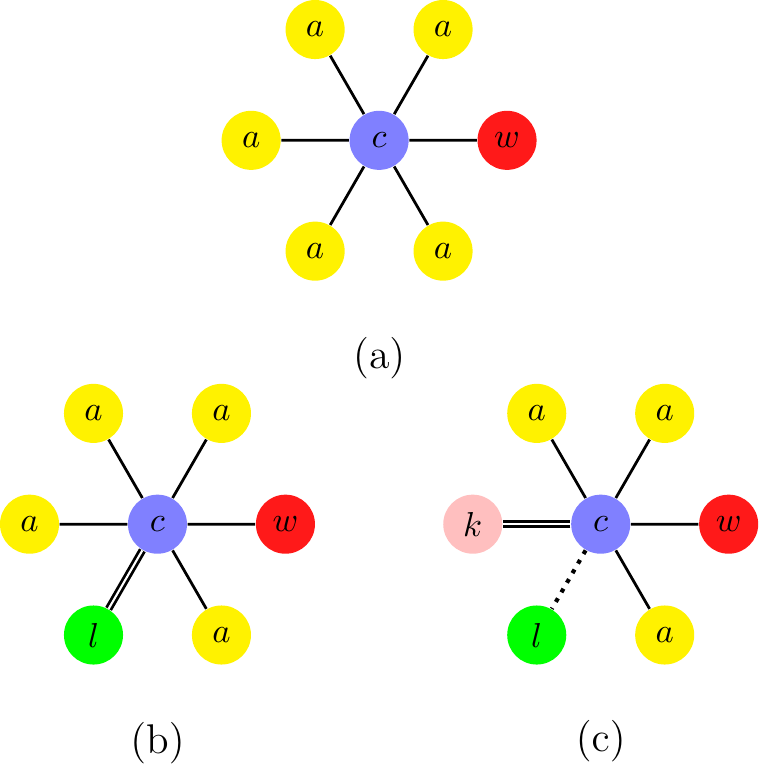}
	\caption{Star $S_7$ of order $7$ for quantum transport. The trap vertex $w$ is colored in red and is one of the outer vertices, not the central one $c$. (a) Unperturbed graph. Engineered graphs to achieve the maximum transport efficiency  (b) for a localized initial state and Hamiltonian \eqref{eq:ls_sg_out_ham} and (c) for an initial superposition of two vertices and Hamiltonian \eqref{eq:ss_sg_out_ham}. The perturbed edges are denoted by doubled and dotted lines. Identically evolving vertices under the given Hamiltonian  are identically colored and labeled.}
	\label{fig:SG_out}
\end{figure}

We now assume that the trap $w$ for quantum transport in $S_N$ is not the central vertex $c$, whose degree is $\deg(c)=N-1$, but one of the outer vertices, say $w$, whose degree is $\deg(w)=1$, Fig\ref{fig:SG_out}(a). In other words, we consider the CBG with $N_1=N-1$ and $N_2=1$.  For clarity, we rewrite the Laplacian matrix of $S_{N}$ as
\begin{equation}
\label{eq:Lapl_sg_outer}
\mathcal{L}=(N-1)\vert c \rangle \langle c \vert+\sum_{j \neq c}\dyad{j} - \sum_{j \neq c}( \ketbra{j}{c} + \ketbra{c}{j})\,.
\end{equation}
The star graph, as well as its Laplacian matrix, is invariant under the permutation of all its outer vertices. Making one of the outer vertices the trap breaks such a symmetry and the resulting transport Hamiltonian \eqref{eq:transport_Ham} is invariant under the permutation of all the outer vertices but the trap. 

The subspace \eqref{eq:Krylov_subspace} for the transport Hamiltonian \eqref{eq:transport_Ham} is three-dimensional and, according to \eqref{eq:basis_cbg}, is spanned by the basis states
\begin{align}
\ket{e_1} &= \ket{w}\,, \nonumber\\
\ket{e_2} &= \ket{c}\,, \nonumber\\
\ket{e_3} &= \dfrac{1}{\sqrt{N-2}} \sum_{j \neq c,w} \ket{j}\,.
\label{eq:basis_sg_out}
\end{align}

According to \eqref{eq:transport_eff_ideal}, the transport efficiency for an initial localized state $\ket{l \neq w}$ is
\begin{equation}
\eta = \begin{dcases}
1 & \text{if $l = c$,}\\
\frac{1}{N-2} & \text{if $l \neq c,w$,}
\end{dcases}
\label{eq:eta_sg__out_l}
\end{equation}
while for any superposition of two vertex states $(\ket{l} + e^{i\gamma}\ket{k})/\sqrt{2}$, we have
\begin{equation}
\eta = \begin{dcases}
\frac{N-1}{2(N-2)} & \text{if $l = c$ or $k=c$,}\\
\frac{1+\cos \gamma}{N-2} & \text{if $l,k \neq c,w$,}
\end{dcases}
\label{eq:eta_sg_out_s}
\end{equation}
The walker is completely absorbed at the trap ($\eta=1$) if initially localized in the central vertex and for $N=4$ and $\gamma = 0$ for the initial superposition of states. In both cases, if the initial state does not involve the central vertex $c$, $\eta \sim 1/N$ for $N\gg 1$ and $\lim_{N \to +\infty} \eta = 0$. Instead, for the superposition of two vertex states of which one is $\ket{c}$, $\eta$ approaches $1/2$ in the limit of $N\to +\infty$.

\subsubsection{Localized Initial State}
To improve the transport efficiency for a localized initial state $\ket{l}$, we perturb the edge $(l,c)$ which connects the starting vertex and the central vertex $c$, not the trap $w$, Fig.\ref{fig:SG_out}(b). The complete Hamiltonian \eqref{eq:engineered_Ham} is therefore
\begin{align}
\label{eq:ls_sg_out_ham}
\mathcal{H}_{c} = \mathcal{H}_{t} + \lambda e^{-i\theta}\ket{l}\bra{c} + \lambda e^{i\theta}\ket{c}\bra{l}. 
\end{align}
The perturbation breaks the symmetry of the star graph with one trap. This is also reflected in the Krylov subspace $\mathcal{I}(\mathcal{H},\ket{w})$ (see Appendix \ref{app:basis_per_sg_out_tra}), which, despite being three-dimensional as that for the unperturbed transport Hamiltonian, is spanned by different  basis states
\begin{align}
\label{eq:basis_sg_out_l}
\ket{e_1} &= \ket{w}\,,\nonumber\\
\ket{e_2} &= \ket{c}\,,\nonumber\\
\ket{e_3} &= \dfrac{-\sum_{j \neq c,w,l} \ket{j} + \left( \lambda e^{-i\theta} - 1 \right) \ket{l}}{\sqrt{N-3 + \left| \lambda e^{-i\theta} -1 \right|^{2}}} \,.
\end{align}
The perturbation alters the Krylov subspace relevant to the computation of the transport efficiency since the three bases for the unperturbed and perturbed system, \eqref{eq:basis_sg_out} and \eqref{eq:basis_sg_out_l}, respectively, span different three-dimensional subspaces.

Therefore, the transport efficiency \eqref{eq:transport_eff_ideal} for the initial localized state $\ket{l}$ is
\begin{align}
\eta(\lambda,\theta) &= \left\vert \braket{e_3}{l}\right\vert^2 = \dfrac{\left| \lambda e^{-i\theta} - 1 \right|^2}{N-3 + \left| \lambda e^{-i\theta} -1 \right|^{2}}\nonumber\\
&=1-\frac{N-3}{\lambda^2-2\lambda \cos(\theta)+N-1}\,.
\label{eq:eta_sg_out_l}
\end{align}
The star graph $S_3$ provides $\eta =1$ independently of the perturbation. For $N>3$, instead, the transport efficiency does depend on the perturbation, both $\lambda$ and $\theta$, and its maximum is
\begin{equation}
    \eta_{max} = \eta(\lambda,\pi) = \frac{(\lambda + 1)^2}{\lambda^2+2\lambda+N-2}
\end{equation}
and, as evident also from \eqref{eq:eta_sg_out_l}, approaches the optimal value 1 in the limit of an infinitely strong perturbation
\begin{equation}
\lim_{\lambda \to +\infty} \eta \left( \lambda ,\theta \right) = 1\,.
\end{equation}
Perturbing an edge $(c,l)$, with $l \neq w$, still makes the initial localized state $\ket{c}$ achieve $\eta = 1$. The transport efficiency of an initial localized state $\ket{a}$ with $a \neq c,l,w$ is
\begin{align}
\eta(\lambda,\theta) &= \left\vert \braket{e_3}{a}\right\vert^2 = \frac{1}{\lambda^2-2\lambda \cos \theta+N-2}\,.
\end{align}


\subsubsection{Superposition State}
To improve the transport efficiency for an initial superposition of two vertex states, $( \ket{l} + e^{i\gamma}\ket{k})/\sqrt{2}$, we perturb the edges $(l,c)$ and $(k,c)$ which connect the two vertices of the initial state and the central vertex $c$, Fig.\ref{fig:SG_out}(c). The complete Hamiltonian \eqref{eq:engineered_Ham} is therefore
\begin{align}
\label{eq:ss_sg_out_ham}
\mathcal{H}_{c} =& \mathcal{H}_{t} + \lambda \ket{l}\bra{c} + \lambda \ket{c}\bra{l}\nonumber \\
&+\lambda e^{-i\theta} \ket{k}\bra{c} + \lambda e^{i\theta} \ket{c}\bra{k}\,.
\end{align}
We consider the phase factor only for one edge to have the minimum number of degrees of freedom.

Similarly to the case of the localized state, the resulting Krylov subspace is three-dimensional and is spanned by the states (see Appendix \ref{app:basis_per_sg_out_tra})
\begin{align}
\label{eq:basis_sg_out_sup}
\ket{e_1} &= \ket{w}\,,\nonumber\\
\ket{e_2} &= \ket{c}\,,\nonumber\\
\ket{e_3} &= \dfrac{-\sum_{j \neq c,w,l,k} \ket{j}  + \left( \lambda -1 \right)\ket{l}+\left( \lambda e^{-i\theta} -1 \right)  \ket{k}}{\sqrt{N-4 + (\lambda - 1 )^{2} +\left| \lambda e^{-i\theta} - 1 \right|^{2}}}\,.
\end{align}
Therefore, the transport efficiency \eqref{eq:transport_eff_ideal} for the initial superposition of two vertex states is 
\begin{align}
\eta(\lambda,\theta,\gamma) &= \left\vert \braket{e_3}{\psi_0}\right\vert^2\nonumber\\
&= \dfrac{\left| \left( \lambda  - 1 \right) + e^{i\gamma}\left( \lambda e^{i\theta} - 1 \right) \right|^2}{2(N-4 +\left| \lambda e^{-i\theta} - 1 \right|^{2}+ (\lambda  - 1)^{2})}.
\end{align}
It depends not only on the perturbation, $\lambda$ and $\theta$, but also on the specific initial superposition, $\gamma$. 

Similarly to the discussion of Sec. \ref{subsec:SG_central_ss}, we observe that $\eta(\lambda,\theta=0, \gamma=0) =1$ for $N=4$, otherwise the transport efficiency is never optimal, $\eta <1$.

\section{Null Transport Efficiency} \label{s:null}
So far, our discussion has been about  achieving unit transport efficiency. 
However, sometimes one might be interested in achieving the opposite result, $\eta = 0$, e.g., when the purpose is to avoid loss phenomena. To obtain a null transport efficiency, the walker must never reach the trap vertex $\ket{w}$, which is where the loss process occurs. For any given initial state $\ket{\psi_0}$, this condition can be written as 
\begin{equation}
\label{eq:null_taylor_vec}
\langle e_{k} \vert \psi_{0} \rangle = 0 \quad \forall k\,,
\end{equation}
where $\{ \ket{e_k}\}_{k = 1,\ldots,m}$ are the basis states of the $m$-dimensional Krylov subspace required to compute the transport efficiency \eqref{eq:transport_eff_ideal}. 
In other words, the walker will never reach the trap vertex $\ket{w}$ only if $\ket{\psi_{0}}$ is orthogonal to that Krylov subspace.

To achieve $\eta=0$, the first obvious condition is $\langle w \vert \psi_{0} \rangle = 0$. The remaining conditions depend on the given graph, i.e., on the \textit{unperturbed} transport Hamiltonian. For the complete graph and the star graph with central trap [basis \eqref{eq:basis_cg}] those conditions read as follows:
\begin{equation}
\begin{cases}
\label{null_cg}
\langle w \vert \psi_{0} \rangle = 0 \,,\\
\sum_{j \neq w} \langle j \vert \psi_{0} \rangle = 0\,,
\end{cases}
\end{equation}
and those for the complete bipartite graph [basis \eqref{eq:basis_cbg}] are
\begin{equation}
\begin{cases}
\label{null_cbg}
\langle w \vert \psi_{0} \rangle = 0 \,,\\
\sum_{j \in V_1 \neq w} \langle j \vert \psi_{0} \rangle = 0\,,\\
\sum_{k \in V_2} \langle k \vert \psi_{0} \rangle = 0\,,
\end{cases}
\end{equation}
and those for the star graph with outer trap [basis \eqref{eq:basis_sg_out}] are
\begin{equation}
\begin{cases}
\label{null_sg_ot}
\langle w \vert \psi_{0} \rangle = 0 \,,\\
\sum_{j \neq w,c} \langle j \vert \psi_{0} \rangle = 0\,,\\
\langle c \vert \psi_{0} \rangle = 0\,,
\end{cases}
\end{equation}
where $c$ denotes the central vertex of the star graph. According to these conditions, it is immediately clear that a localized state will always provide a nonzero transport efficiency, since a localized state has always nonzero overlap with at least one of the basis states.

Let us now discuss some relations between stationary states of the \textit{unperturbed} transport Hamiltonian \eqref{eq:transport_Ham} and states providing $\eta = 0$ for the graphs considered. Any state $\ket{\nu}$ that satisfies \eqref{null_cg} is also a stationary state of the transport Hamiltonian of the complete graph, as well as of the Laplacian matrix,
\begin{align}
\label{eigenstate_cg}
\mathcal{H}_{t}^{(cg)} \ket{\nu} = \mathcal{L}^{(cg)}\ket{\nu} =  N \ket{\nu} - \sum_{j,h} \langle j \vert \nu \rangle \ket{h} = N \ket{\nu}\,.
\end{align}
The same happens for the star graph with central trap, where any state $\ket{\mu}$ that satisfies \eqref{null_cg} is also a stationary state of the transport Hamiltonian, as well as of the Laplacian matrix,
\begin{align}
\label{eigenstate_sg}
\mathcal{H}_{t}^{(sg)} \ket{\mu} = \mathcal{L}^{(sg)}\ket{\mu} =  \ket{\mu} - \sum_{j} \langle j \vert \mu \rangle \ket{w} = \ket{\mu}\,.
\end{align}
Instead, in the complete bipartite graph, as long as the parameters $N_1$ and $N_2$ are different, a state $\ket{\zeta}$ that satisfies \eqref{null_cbg} may not be a stationary state. As an example, the state
\begin{align}
\label{null_not_stat}
\ket{\zeta} &= \dfrac{1}{2} \left[ \left( \ket{l} -\ket{k}\right) + \left( \ket{r} -\ket{s}\right) \right]\,,
\end{align}
with $l,k \in V_1 \setminus \{w\}$ and $r,s \in V_2$, fulfills the condition \eqref{null_cbg}, but it is not an eigenstate of the Laplacian or transport Hamiltonian of the CBG, since
\begin{equation}
\mathcal{H}_{t}^{(cbg)}\ket{\zeta} = \dfrac{1}{2} \left[ N_2\left( \ket{l} -\ket{k}\right) + N_1\left( \ket{r} -\ket{s}\right) \right] \not\propto \ket{\zeta}\,.
\end{equation}
As a final remark, we point out that any eigenstate of the transport Hamiltonian $\mathcal{H}_{t}$  \eqref{eq:transport_Ham} associated to a real eigenvalue\footnote{Eigenstates associated to complex eigenvalues $\varepsilon$ decay or increase with time, depending on the sign of $ \varepsilon $. Here the total density is subjected to the loss process since the imaginary part of the sum of the eigenvalues of $\mathcal{H}_t$ is  $\Im [\Tr (\mathcal{H}_t)] = -\kappa < 0$.} is a stationary state and provides null transport efficiency. Instead, a state with null transport efficiency is not necessarily a stationary state. Being a stationary state is a sufficient, but not necessary, condition for having $\eta = 0$. As an example, any linear combination of the Hamiltonian eigenstates with real eigenvalues is not a stationary state, but it still provides $\eta = 0$.

\section{Conclusions}
\label{conclusion}
In conclusion, we have investigated in details the transport efficiency of a quantum walker on highly symmetric graphs. In particular, we have analyzed situations where the walker is initially localized at a vertex, or may be prepared in a superposition of two vertex states, and the goal of the transport protocol is to trap the walker at a specific trap vertex. In our scheme there is only one trap vertex, $w$, which is accountable for the loss processes with a rate $\kappa$. The (transport) Hamiltonian of the system is $H_t = \mathcal{L}-i \kappa \vert w \rangle \langle w \vert$, where $\mathcal{L}$ is the Laplacian matrix of the graph. We have analyzed in detail the dynamics of CTQWs on complete, complete bipartite, and star graph.

We have analytically evaluated the transport efficiency by computing the overlap of the initial state with the Krylov subspace $\mathcal{I}$ relevant to dynamics of the system. After that, we have engineered the graph with the minimal perturbation required to achieve the optimal transport efficiency, $\eta = 1$. The perturbation consists of adding an extra weight in the form $\lambda e^{i \theta}$, with $\lambda > 0$ and $0\leq \theta < 2\pi$, to one or two edges for the initial localized state, $\ket{l}$, or the superposition of two vertex states, $(\ket{l}+e^{i\gamma}\ket{k})/\sqrt{2}$, respectively. The perturbed edges involve the vertices used to define the initial state and the perturbation induces chirality in the system. 

We have started our analysis by considering CTQWs on unperturbed complete and bipartite complete graphs, since they cannot achieve unit transport efficiency for initially localized states. We prove that introducing a perturbation breaks the symmetries of the original graph, increases the dimension of the Krylov subspace, and allows one to achieve unit transport efficiency. This is true also for an initial superposition of two vertex states, and in this case the two edges should be perturbed by terms with matching phases.  We have also considered the case of the star graph, where the minimal perturbation mentioned above is not sufficient to achieve unit transport efficiency. Indeed, the perturbation affects the Krylov subspace but preserves its dimension with respect to the unperturbed case. As a result, the initial state may have a component outside of the Krylov subspace and the transport efficiency may not achieve unit value. 
On the other hand, if the trap is one of the outer vertices, the initial state localized at the central vertex leads to unit transport efficiency even for the unperturbed graph.

Overall, our results suggest that breaking the symmetries of a graph may improve the transport efficiency. In particular, we have shown that unit {transport} efficiency is possible when a graph may be perturbed in such a way that the dimension of the relevant Krylov subspace is increased. 
In general, the needed perturbation corresponds to the introduction of a slowly varying external magnetic field, which adds Peierls' phases to the system \cite{PhysRevA.101.032336,fri21} and, in turn, induces chirality. Experimentally, directional continuous-time  quantum walks can be realized using nuclear magnetic resonance \cite{PhysRevA.93.042302} or linear optics \cite{Wang20}.
Our results also show that unit transport efficiency is possible without environmental noise, and pave the way to graph engineering for enhanced quantum transport.

\begin{acknowledgments}
Work was done under the auspices of GNFM-INdAM.
\end{acknowledgments}

\appendix

\section{Basis of the Krylov subspace related to the unperturbed transport Hamiltonian}
In this appendix we show how to construct by means of the dimensionality reduction method (Sec. \ref{sec:QWs}) the basis states of the Krylov subspace $\mathcal{I}(\mathcal{H},\ket{w})$ \eqref{eq:Krylov_subspace} required to compute the transport efficiency \eqref{eq:transport_eff_ideal} for the unperturbed transport Hamiltonian $\mathcal{H}$ \eqref{eq:transport_Ham} of each graph discussed. The first basis state is always the trap, $\ket{e_1}=\ket{w}$. In the following, O.N. denotes the orthonormalization of the current state $\mathcal{H}\ket{e_k}$ with respect to the previous basis states $\{\ket{e_1},\ldots,\ket{e_k}\}$ and, for brevity, we denote the Krylov subspace by $\mathcal{I}$. The procedure stops when we find the minimum index $m$ such that $\mathcal{H} \ket{e_m} \in \operatorname{span}(\left\lbrace \ket{e_1},\ldots,\ket{e_m}\right\rbrace)$, as this then implies  $\mathcal{H}^k \ket{w} \in \operatorname{span}(\left\lbrace \ket{e_1}, \ldots, \ket{e_m} \right\rbrace)=:\mathcal{I}$ $\forall k \in \mathbb{N}_0$.

\subsection{Complete graph}
\label{app:basis_cg}
The Laplacian matrix of the complete graph $K_N$ is \eqref{eq:Lapl_cg}, thus the unperturbed transport Hamiltonian is 
\begin{equation}
\mathcal{H}=N\sum_{j=1}^{N} \ket{j}\bra{j} - \sum_{j,h=1}^{N} \ket{j}\bra{h} -i\kappa \ket{w}\bra{w}\,.
\end{equation}
The basis states \eqref{eq:basis_cg} are obtained as follows:
\begin{align}
\mathcal{H}\ket{e_1} &= (N-1-i\kappa) \ket{w} -\sum_{j \neq w}^{N} \ket{j}\nonumber\\ 
&=(N-1-i\kappa)\ket{e_1}-\sqrt{N-1}\ket{e_2}  \nonumber \\
&\xrightarrow[]{\text{O.N.}} \ket{e_2}\,, \\
\mathcal{H} \ket{e_2} &\in \operatorname{span}(\{\ket{e_1},\ket{e_2}\})\, \nonumber .
\end{align}

\subsection{Complete bipartite graph}
\label{app:basis_cbg}
These calculations are given in \cite{razzoli2021transport} and are reported here for completeness. The Laplacian matrix of the CBG graph $K_{N_1,N_2}$ is \eqref{eq:Lapl_cbg}, thus the unperturbed transport Hamiltonian is
\begin{align}
\mathcal{H}=&N_2 \sum_{j \in V_1} \vert j \rangle \langle j \vert +N_1 \sum_{h \in V_2} \vert h \rangle \langle h \vert \nonumber\\
&-\sum_{j \in V_1}\sum_{h \in V_2}(\vert j \rangle \langle h \vert +\vert h \rangle \langle j \vert )-i\kappa \vert w \rangle \langle w \vert \,.
\end{align}
Recalling that we assume $w \in V_1$, the basis states \eqref{eq:basis_cbg} are obtained as follows:
\begin{align}
\mathcal{H}\ket{e_1} &= (N_2-i\kappa) \ket{w} -\sum_{h \in V_2} \ket{h} \nonumber\\
&=(N_2-i\kappa)\ket{e_1}-\sqrt{N_2}\ket{e_2} \xrightarrow[]{\text{O.N.}} \ket{e_2}\,, \\
\mathcal{H}\ket{e_2} &= \frac{N_1}{\sqrt{N_2}} \sum_{h \in V_2} \ket{h} -\frac{1}{\sqrt{N_2}}\sum_{j \in V_1}\sum_{h \in V_2} \ket{j}\nonumber\\
&=N_1 \ket{e_2}-\sqrt{N_2}\sum_{j \in V_1 \setminus \{w\}}\ket{j} - \sqrt{N_2} \ket{e_1}\nonumber \nonumber  \\
&= N_1 \ket{e_2} - \sqrt{N_2(N_1-1)}\ket{e_3}-\sqrt{N_2}\ket{e_1} \nonumber \\ &\xrightarrow[]{\text{O.N.}} \ket{e_3}\,, \\
\mathcal{H}\ket{e_3} &\in \operatorname{span}(\{\ket{e_1},\ket{e_2},\ket{e_3}\})\, \nonumber .
\end{align}

\subsection{Star graph: Central Trap}
\label{app:basis_sg}
The Laplacian matrix of the star graph $S_N$ is \eqref{eq:Lapl_sg_central}, thus the unperturbed transport Hamiltonian is
\begin{align}
\mathcal{H}&=(N-1) \vert w \rangle \langle w \vert - \sum_{j \neq w}^{N} \left( \vert w \rangle \bra{j} + \ket{j} \langle w \vert - \ket{j} \bra{j} \right) \nonumber\\
&-i\kappa \ket{w}\bra{w}\,,
\end{align}
where $w$ is the central vertex. The basis states are the same as those of the complete graph \eqref{eq:basis_cg} and are obtained as follows
\begin{align}
\mathcal{H}\ket{e_1} &= (N-1-i\kappa) \ket{w} -\sum_{j \neq w}^{N} \ket{j} \nonumber\\
&=(N-1-i\kappa)\ket{e_1}-\sqrt{N-1}\ket{e_2} \nonumber \\ &\xrightarrow[]{\text{O.N.}} \ket{e_2}\,, \\
\mathcal{H}\ket{e_2} &\in \operatorname{span}(\{\ket{e_1},\ket{e_2}\})\, \nonumber .
\end{align}

\section{Basis of the Krylov subspace related to the Hamiltonian engineered for maximum transport efficiency}
In this appendix we show how to construct by means of the dimensionality reduction method (Sec. \ref{sec:QWs}) the basis states of the Krylov subspace $\mathcal{I}(\mathcal{H},\ket{w})$ \eqref{eq:Krylov_subspace} required to compute the transport efficiency \eqref{eq:transport_eff_ideal} for the perturbed transport Hamiltonian $\mathcal{H}$ \eqref{eq:engineered_Ham} of each graph discussed. The Hamiltonian is properly defined and perturbed depending on the initial state considered, localized at a vertex, $\ket{\psi_{0}}=\ket{l}$, or superposition of two vertex states, $\ket{\psi_{0}}=\left( \ket{l} + e^{i\gamma}\ket{k} \right)/\sqrt{2}$, with $l,k \neq w$ (vertices different from the trap). The first basis state is always the trap, $\ket{e_1}=\ket{w}$. In the following, O.N. denotes the orthonormalization of the current state $\mathcal{H}\ket{e_k}$ with respect to the previous basis states $\{\ket{e_1},\ldots,\ket{e_k}\}$ and, for brevity, we denote the Krylov subspace by $\mathcal{I}$. To simplify the calculation, we neglect the normalization constant of the state $\ket{e_n}$ when computing $\mathcal{H}\ket{e_n}$. Indeed, the normalization constant contributes only as a multiplicative constant, which can be absorbed in the final normalization procedure to get $\ket{e_{n+1}}$. The procedure stops when we find the minimum index $m$ such that $\mathcal{H} \ket{e_m} \in \operatorname{span}(\left\lbrace \ket{e_1},\ldots,\ket{e_m}\right\rbrace)$, as this then implies  $\mathcal{H}^k \ket{w} \in \operatorname{span}(\left\lbrace \ket{e_1}, \ldots, \ket{e_m} \right\rbrace)=:\mathcal{I}$ $\forall k \in \mathbb{N}_0$. 

\subsection{Complete Graph}
\label{app:basis_per_cg}
The Laplacian matrix of the complete graph $K_N$ is \eqref{eq:Lapl_cg}.

\subsubsection{Localized State}
The complete Hamiltonian to achieve the optimal transport efficiency for the initial localized state $\ket{l}$ is
\begin{align}
\mathcal{H}=&N\sum_{j=1}^{N} \vert j \rangle \langle j \vert - \sum_{j,h=1}^{N} \vert j \rangle \langle h \vert -i \kappa \vert w \rangle \langle w \vert \nonumber\\
&+ \lambda e^{i\theta} \vert w\rangle \bra{l} + \lambda e^{-i\theta} \ket{l} \bra{w} \,.
\end{align}
The basis states \eqref{eq:basis_per_cg} are obtained as follows:
\begin{align}
\mathcal{H}\ket{e_1} &= (N-1-i\kappa) \ket{w} -\sum_{j \neq w}^{N} \ket{j} +\lambda e^{-i\theta} \ket{l} \nonumber\\
&  \xrightarrow[]{\text{O.N.}} \ket{e_2}\,, \\
\mathcal{H}\ket{e_2} &\propto \lambda \left( \lambda - e^{i\theta} \right) \ket{w} + \left( N-1-\lambda e^{-i\theta} \right) \sum_{j}^{N} \ket{j} \nonumber \\
&\quad+ N \ket{e_2} \xrightarrow[]{\text{O.N.}} \ket{e_3}\,, \\
\mathcal{H}\ket{e_3} &\in \operatorname{span}(\{\ket{e_1},\ket{e_2},\ket{e_3}\})\, \nonumber .
\end{align}

\subsubsection{Superposition State}
The complete Hamiltonian to achieve the optimal transport efficiency for the initial superposition of two vertex states is
\begin{align}
\mathcal{H}=&N\sum_{j=1}^{N} \vert j \rangle \langle j \vert - \sum_{j,h=1}^{N} \vert j \rangle \langle h \vert -i \kappa \vert w \rangle \langle w \vert +\lambda  \vert w \rangle \langle l \vert \nonumber \\
&+ \lambda  \vert l \rangle \langle w \vert + \lambda e^{i\theta} \vert w\rangle \langle k \vert + \lambda e^{-i\theta} \vert k \rangle \langle w \vert \,.
\end{align}
The basis states \eqref{eq:basis_sup_per_cg} are obtained as follows:
\begin{align}
\mathcal{H}\ket{e_1} &= (N-1-i\kappa) \ket{w} -\sum_{j \neq w}^{N} \ket{j} +\lambda e^{-i\theta} \ket{k} + \lambda  \ket{l} \nonumber\\
& \xrightarrow[]{\text{O.N.}} \ket{e_2} \,, \\
\mathcal{H}\ket{e_2} &\propto N \ket{e_2} - \left[ -(N-1) + \lambda (1+e^{-i\theta}) \right] \sum_{j}^{N} \ket{j} \nonumber \\
&\quad+ \left[ \lambda (\lambda  - 1) + \lambda e^{i\theta}(\lambda e^{-i\theta} - 1) \right]\ket{w} \nonumber \\
&\xrightarrow[]{\text{O.N.}} \ket{e_3}\,, \\
\mathcal{H}\ket{e_3} &\in \operatorname{span}(\{\ket{e_1},\ket{e_2},\ket{e_3}\})\, \nonumber .
\end{align}

\subsection{Complete bipartite graph}
The Laplacian matrix of the CBG  $K_{N_1,N_2}$ is \eqref{eq:Lapl_cbg}.

\label{app:basis_per_cbg}
\subsubsection{Localized State}
In the CBG there can be two different initial localized states $\ket{l}$, i) $l \in V_2$ and ii) $l\in V_1$ (Fig. \ref{fig:CBG}), thus we consider two different Hamiltonian to achieve the optimal transport efficiency. In the first case the complete Hamiltonian is
\begin{align}
\mathcal{H}=&N_2 \sum_{j \in V_1} \vert j \rangle \langle j \vert +N_1 \sum_{h \in V_2} \vert h \rangle \langle h \vert-\sum_{j \in V_1}\sum_{h \in V_2}(\vert j \rangle \langle h \vert \nonumber \\ 
&+\vert h \rangle \langle j \vert) -i\kappa \vert w \rangle \langle w \vert +\lambda e^{i\theta}\vert w \rangle \langle l \vert + \lambda e^{-i\theta}\vert l \rangle \langle w \vert  \,.
\end{align}
The basis states \eqref{eq:basis_per_cbg} are obtained as follows:
\begin{align}
\mathcal{H}\ket{e_1} &= (N_2-i\kappa) \ket{w} -\sum_{k \in V_2} \ket{k} +\lambda e^{-i\theta} \ket{l} \nonumber \\
&\xrightarrow[]{\text{O.N.}} \ket{e_2} \,,\\ 
\mathcal{H}\ket{e_2} &\propto N_1 \ket{e_2} + \left( N_2 - \lambda e^{-i\theta}  \right) \sum_{j \in V_1} \ket{j}  \nonumber \\
&\xrightarrow[]{\text{O.N.}} \ket{e_3} \,,\\
\mathcal{H}\ket{e_3} &\propto N_2 \ket{e_3} - \left( N_1 -1 \right) \sum_{h \in V_2} \ket{h} \xrightarrow[]{\text{O.N.}} \ket{e_4} \,,\\
\mathcal{H}\ket{e_4} &\in \operatorname{span}(\{\ket{e_1},\ket{e_2},\ket{e_3},\ket{e_4}\}) \,.\nonumber
\end{align}

In the second case the complete Hamiltonian is
\begin{align}
\label{ham_cbg_loc_sup}
\mathcal{H}=&N_2 \sum_{j \in V_1} \vert j \rangle \langle j \vert +N_1 \sum_{h \in V_2} \vert h \rangle \langle h \vert -\sum_{j \in V_1}\sum_{h \in V_2}(\vert j \rangle \langle h \vert \nonumber \\
&+\vert h \rangle \langle j \vert) -i\kappa \vert w \rangle \langle w \vert +\lambda e^{i\theta}\vert m \rangle \langle l \vert + \lambda e^{-i\theta}\vert l \rangle \langle m \vert\,,  
\end{align}
and the basis states \eqref{eq:basis_per_cbg_2nn} are obtained as follows:
\begin{align}
\mathcal{H}\ket{e_1} &= (N_2-i\kappa) \ket{w} -\sum_{h \in V_2} \ket{h} \xrightarrow[]{\text{O.N.}} \ket{e_2}\,, \\ 
\mathcal{H}\ket{e_2} &\propto -N_{2} \sum_{j \in V_1} \ket{j} + \lambda e^{-i\theta} \ket{l} \xrightarrow[]{\text{O.N.}} \ket{e_3} \,,\\
\mathcal{H}\ket{e_3} &\propto N_2 \ket{e_3} - \left[ N_2 \left( N_1 -2 \right) + \left( \lambda e^{i\theta} -N_2 \right) \right] \sum_{h \in V_2} \ket{h} \nonumber \\
& \quad+ \lambda e^{-i\theta} \left( \lambda e^{i\theta} - N_2 \right) \ket{m} \xrightarrow[]{\text{O.N.}} \ket{e_4}  \,,\\
\mathcal{H}\ket{e_4} &\propto N_1 \ket{e_4} + \left(1-\dfrac{1}{N_2} \right) \lambda e^{i\theta} \ket{m} \xrightarrow[]{\text{O.N.}} \ket{e_5} \,,\\
\mathcal{H}\ket{e_5} &\in \operatorname{span}(\{\ket{e_1},\ket{e_2},\ket{e_3},\ket{e_4},\ket{e_5}\}) \,.\nonumber
\end{align}


\subsubsection{Superposition State}
There are three different initial states to be considered: (1) $l,k \in V_2$, (2) $l \in V_1 \setminus \{w\}$ and $k \in V_1$, (3) $l\in V_1$ and $k \in V_2$. For the case iii) the Hamiltonian \eqref{ham_cbg_loc_sup} already leads to perfect quantum transport, thus we will focus on the two remaining cases.

In the first case the complete Hamiltonian is
\begin{align}
\mathcal{H}=&N_2 \sum_{j \in V_1} \vert j \rangle \langle j \vert +N_1 \sum_{h \in V_2} \vert h \rangle \langle h \vert -\sum_{j \in V_1}\sum_{h \in V_2}(\vert j \rangle \langle h \vert \nonumber \\
&+\vert h \rangle \langle j \vert) -i\kappa \vert w \rangle \langle w \vert +\lambda e^{i\theta}\ket{w} \bra{k} + \lambda e^{-i\theta}\ket{k} \bra{w} \nonumber \\
& +\lambda \ket{w} \bra{l} + \lambda \ket{l}\bra{w}\,,
\end{align}
and the basis states \eqref{eq:basis_sup_per_cbg} are obtained as follows:
\begin{align}
\mathcal{H}\ket{e_1} &= (N_2-i\kappa) \ket{w} -\sum_{h \in V_2} \ket{h} +\lambda e^{-i\theta} \ket{k} + \lambda \ket{l}\nonumber \\
& \xrightarrow[]{\text{O.N.}} \ket{e_2} \,,\\
\mathcal{H} \ket{e_2} &\propto \left[ N_2 -1 - \left( \lambda e^{-i\theta} - 1 \right) + \left( \lambda -1 \right) \right] \sum_{j \in V_1} \ket{j} \nonumber \\
& \quad+N_1 \ket{e_2}  \xrightarrow[]{\text{O.N.}} \ket{e_3} \,,\\
\mathcal{H}\ket{e_3} &\propto  N_2\ket{e_3} - \left( N_1 -1 \right) \sum_{h \in V_2} \ket{h} \xrightarrow[]{\text{O.N.}} \ket{e_4} \,,\\
\mathcal{H}\ket{e_4} &\in \operatorname{span}(\{\ket{e_1},\ket{e_2},\ket{e_3},\ket{e_4}\}) \,.\nonumber
\end{align}

In the second case the complete Hamiltonian is
\begin{align}
\mathcal{H}=&N_2 \sum_{j \in V_1} \vert j \rangle \langle j \vert +N_1 \sum_{h \in V_2} \vert h \rangle \langle h \vert -\sum_{j \in V_1}\sum_{h \in V_2}(\vert j \rangle \langle h \vert \nonumber \\
&+ \vert h \rangle \langle j \vert) -i\kappa \vert w \rangle \langle w \vert+\lambda e^{i\theta}\ket{w} \bra{m} + \lambda e^{-i\theta}\ket{k} \bra{m} \nonumber \\
&+ \lambda \ket{m} \bra{l}  + \lambda \ket{l}\bra{m} \,,
\end{align}
and the basis states \eqref{eq:basis_sup2_per_cbg} are obtained as follows: 
\begin{align}
\mathcal{H}\ket{e_1} &= (N_2-i\kappa) \ket{w} -\sum_{h \in V_2} \ket{h}
\xrightarrow[]{\text{O.N.}} \ket{e_2}\,,\\
\mathcal{H}\ket{e_2} &\propto N_1 \ket{e_2} - N_2 \sum_{j \in V_1} \ket{j} + \lambda e^{-i\theta} \ket{k} + \lambda \ket{l}  \nonumber \\
&\xrightarrow[]{\text{O.N.}} \ket{e_3} \,,\\
\mathcal{H}\ket{e_3} &\propto \left[ N_2\left(N_1-1 \right) - \lambda -  \lambda e^{-i\theta} \right] \sum_{h \in V_2} \ket{h} \nonumber \\
&\quad + \left[ \lambda \left( \lambda - N_2 \right) + \lambda e^{i\theta} \left( \lambda e^{-i\theta} - N_2 \right) \right] \ket{m} \nonumber \\
&\xrightarrow[]{\text{O.N.}} \ket{e_4} \,,\\
\mathcal{H}\ket{e_4} &\propto \left[ \left( 1 - \dfrac{1}{N_2} \right) +  N_2 - 1  \right] \sum_{j \in V_1} \ket{j} +\lambda \left( 1 - \dfrac{1}{N_2} \right) \ket{l}\nonumber \\
& \quad  + \lambda e^{-i\theta} \left( 1 - \dfrac{1}{N_2} \right) \ket{k} \xrightarrow[]{\text{O.N.}} \ket{e_5} \,,\\
\mathcal{H}\ket{e_5} &\in \operatorname{span}(\{\ket{e_1},\ket{e_2},\ket{e_3},\ket{e_4},\ket{e_5}\})\,. \nonumber
\end{align}

\subsection{Star graph: Central trap}
\label{app:basis_per_sg_cen_tra}
The Laplacian matrix of the star graph  $S_{N}$ is \eqref{eq:Lapl_sg_central}. Since the trap vertex is the central one, the initial states can involve only the outer vertices (Fig. \ref{fig:SG}).


\subsubsection{Localized State}
The complete Hamiltonian for the initial localized state $\ket{l}$ is

\begin{align}
\mathcal{H}=&(N-1) \vert w \rangle \langle w \vert - \sum_{j \neq w}^N \left( \vert w \rangle \langle j \vert + \vert j \rangle \langle w \vert+\vert j \rangle \langle j \vert \right) \nonumber\\
&- i \kappa \vert w \rangle \langle w \vert +\lambda e^{i\theta} \vert w\rangle \bra{l} + \lambda e^{-i\theta} \ket{l} \bra{w}\,.
\end{align}

The basis states are the same as those of the complete graph \eqref{eq:basis_cg} and are obtained as follows

\begin{align}
\mathcal{H}\ket{e_1} &= (N-1-i\kappa) \ket{w} -\sum_{j \neq w}^{N} \ket{j} +\lambda e^{-i\theta} \ket{l} \nonumber \\
&\xrightarrow[]{\text{O.N.}} \ket{e_2} \,,\\
\mathcal{H}\ket{e_2} &\in \operatorname{span}(\{\ket{e_1},\ket{e_2}\}) \,. \nonumber
\end{align}

\subsubsection{Superposition State}
The complete Hamiltonian for the initial superposition of two vertex states is
\begin{align}
\mathcal{H}=&(N-1) \vert w \rangle \langle w \vert - \sum_{ j \neq w}^{N} \left( \vert w \rangle \langle j \vert + \vert j \rangle \langle w \vert+\vert j \rangle \langle j \vert \right) \nonumber \\
&- i \kappa \vert w \rangle \langle w \vert + \lambda e^{i\theta} \vert w\rangle \bra{k} + \lambda e^{-i\theta} \ket{k} \bra{w} \nonumber \\
&+ \lambda  \vert w\rangle \bra{l} + \lambda  \ket{l} \bra{w}\,.
\end{align}

The basis states \eqref{eq:basis_sup_sg_central} are obtained as follows

\begin{align}
\mathcal{H}&\ket{e_1} = (N-1-i\kappa) \ket{w} -\sum_{j \neq w}^{N} \ket{j} +\lambda e^{-i\theta} \ket{k} + \lambda  \ket{l}\,, \nonumber \\
&\xrightarrow[]{\text{O.N.}} \ket{e_2}  \\
\mathcal{H}&\ket{e_2} \in \operatorname{span}(\{\ket{e_1},\ket{e_2}\}) \,. \nonumber
\end{align}

\subsection{Star graph: Outer trap}
\label{app:basis_per_sg_out_tra}
We now consider as a trap vertex one of the outer vertices, thus $w \neq c$.

\subsubsection{Localized State}
The complete Hamiltonian for the initial localized state $\ket{l}$, with $l \neq c$, is

\begin{align}
\mathcal{H}=&(N-1) \vert c \rangle \langle c \vert - \sum_{j \neq c}^{N} \left( \vert c \rangle \langle j \vert + \vert j \rangle \langle c \vert+\vert j \rangle \langle j \vert \right) \nonumber\\
&- i \kappa \vert c \rangle \langle c \vert +\lambda e^{i\theta} \vert c \rangle \langle l \vert + \lambda e^{-i\theta} \vert l \rangle \langle c \vert \,. 
\end{align}

The basis states \eqref{eq:basis_sg_out_l} are obtained as follows:

\begin{align}
\mathcal{H}\ket{e_1} &= \ket{c} + \left( 1 -i\kappa \right) \ket{w} \xrightarrow[]{\text{O.N.}} \ket{e_2} \,,
\end{align}

\begin{align}
\mathcal{H}\ket{e_2} &\propto \left( N-1 \right) \ket{e_2} + \ket{w} - \sum_{j \neq w,c}^{N} \ket{j} + \lambda e^{-i\theta} \ket{l} \nonumber \\
&\xrightarrow[]{\text{O.N.}} \ket{e_3} \,,\\
\mathcal{H}\ket{e_3} &\in \operatorname{span}(\{\ket{e_1},\ket{e_2},\ket{e_3}\}) \,. \nonumber
\end{align}

\subsubsection{Superposition State}
The complete Hamiltonian for the initial superposition of two vertex states is

\begin{align}
\mathcal{H}=&(N-1) \vert c \rangle \langle c \vert - \sum_{ j \neq c}^{N} \left( \vert c \rangle \langle j \vert + \vert j \rangle \langle c \vert+\vert j \rangle \langle j \vert \right) \nonumber \\
&- i \kappa \vert w \rangle \langle w \vert + \lambda e^{i\theta} \vert c \rangle \langle k \vert + \lambda e^{-i\theta} \vert k \rangle \langle c \vert \nonumber \\
&+ \lambda  \vert c \rangle \langle l \vert + \lambda  \vert l \rangle \langle c \vert \,.
\end{align}

The basis states \eqref{eq:basis_sg_out_sup} are obtained as follows:

\begin{align}
\mathcal{H}\ket{e_1} &= \ket{c} + \left( 1 -i\kappa \right) \ket{w} \xrightarrow[]{\text{O.N.}} \ket{e_2} \,,\\
\mathcal{H}\ket{e_2} &\propto \left( N-1 \right) \ket{e_2} + \ket{w} - \sum_{j \neq w,c}^{N} \ket{j} + \lambda \ket{l} \lambda e^{-i\theta} \ket{k} \nonumber \\
&\xrightarrow[]{\text{O.N.}} \ket{e_3} \,,\\
\mathcal{H}\ket{e_3} &\in \operatorname{span}(\{\ket{e_1},\ket{e_2},\ket{e_3}\}) \,. \nonumber
\end{align}

\bibliography{Pgautewen.bib}

\end{document}